\documentclass{aa}
\pdfoutput=1
\usepackage{natbib,psfig,graphicx,ulem}
\usepackage{txfonts}
\usepackage{color,soul}

\def\mathnew{\mathsurround=0pt}
\def\simov#1#2{\lower .5pt\vbox{\baselineskip0pt \lineskip-.5pt
\ialign{$\mathnew#1\hfil##\hfil$\crcr#2\crcr\sim\crcr}}}
\def\simgreat{\mathrel{\mathpalette\simov >}}

\def\arcmin{\hbox{$^\prime$}}

\def\arcsect\!{\hbox{$^{\prime\prime}$}}
\def\etal{{\it et~al.\/}}

\def\MeV{Me\kern-0.11em V}
\def\keV{ke\kern-0.11em V}

\def\cha{{\it Chandra\/}}
\def\xmm{{\it XMM-Newton\/}}

\def\ecmss{ergs cm$^{-2}$ s$^{-1}$}

\def\etal{{\it et~al.\/}}
\begin{document}
\title{An extension of the SHARC survey 
\thanks{Based on observations obtained at the Canada-France-Hawaii
Telescope (CFHT), Gemini Observatory, Observatoire de Haute Provence,
German-Spanish Astronomical Center Calar Alto, ESO at La Silla and
Paranal, Apache Point Observatory 3.5-meter telescope and on data
taken from the SDSS, SIMBAD and NASA/IPAC Extragalactic (NED)
Databases (see acknowledgments).}
}

\author{C. Adami\inst{1} \and
M.P. Ulmer\inst{2} \and
F. Durret\inst{3,4} \and
G. Covone\inst{5} \and
E. Cypriano\inst{6} \and
B.P.~Holden\inst{7} \and
R.~Kron\inst{8} \and
G.B.~Lima Neto\inst{9} \and
A.K.~Romer\inst{10} \and
D.~Russeil\inst{1} \and
B.~Wilhite\inst{11}
 }

\offprints{C. Adami \email{christophe.adami@oamp.fr}}

\institute{
LAM, Traverse du Siphon, 13012 Marseille, France
\and
Department Physics \& Astronomy, Northwestern University,
Evanston, IL 60208-2900, USA
\and
Institut d'Astrophysique de Paris, CNRS, UMR~7095, Universit\'e Pierre et 
Marie Curie, 98bis Bd Arago, 75014 Paris, France
\and
Observatoire de Paris, LERMA, 61 Av. de l'Observatoire, 75014 Paris, France
\and
INAF -- Osservatorio Astronomico di Capodimonte, Naples, Italy
\and
Department of Physics and Astronomy, University College London, London WC1E
6BT, UK
\and
UCO/Lick Observatory, University of California, Santa Cruz, CA 95064, USA
\and
University of Chicago, Department of Astronomy and Astrophysics, 5640 South
Ellis Avenue, Chicago, IL 60637, USA
\and
Instituto de Astronomia, Geof\'{\i}sica e C. Atmosf./USP, R. do Mat\~ao 1226, 
05508-090 S\~ao Paulo/SP, Brazil
\and
Astronomy Centre, University of Sussex, Falmer, Brighton BN1 9QJ, UK
\and
Department of Astronomy, National Center for Supercomputing Applications,
University of Illinois, Urbana-Champaign, 1002 W. Green, Urbana, IL 61801, USA
}

\date{Accepted . Received ; Draft printed: \today}

\authorrunning{Adami et al.}

\titlerunning{An extended SHARC survey}

\abstract
{}
{We report on our search for distant clusters of galaxies based on
optical and X-ray follow up observations of X-ray candidates from the SHARC
survey. Based on the assumption that the absence of bright optical
or radio counterparts to possibly extended X-ray sources could be
distant clusters.}
{We have obtained deep optical images and redshifts for several of
these objects and analyzed archive XMM-Newton or Chandra data where
applicable.}
{In our list of candidate clusters, two are probably galaxy structures at
redshifts of z$\sim$0.51 and 0.28. Seven other structures are possibly
galaxy clusters between z$\sim$0.3 and 1. Three sources are identified
with QSOs and are thus likely to be X-ray point sources, and six more
also probably fall in this category.  One X-ray source is spurious or
variable. For 17 other sources, the data are too sparse at this time
to put forward any hypothesis on their nature. We also serendipitously
detected a cluster at z=0.53 and another galaxy concentration which is
probably a structure with a redshift in the [0.15-0.6] range.}
{We discuss these results within the context of future space missions to demonstrate the
  necessity of a wide field of view telescope optimized for the 0.5-2 keV range.}


\maketitle

\section{Introduction}\label{sec:intro}
\sethlcolor{yellow} \setstcolor{red} Clusters of galaxies are
currently used as a complementary tool to WMAP and distant supernovae
to constrain cosmological parameters as well as the
equation of state of dark energy (e.g. Romer et al. 2001, Allen et al.
2004). However, without strong constraints on cluster formation and
evolution, the reliability of clusters as cosmological probes will
remain in doubt.  In particular, their formation redshift is still not
well established (e.g.  Andreon et al.  2004, Holden et
al. 2004). This is one of the major quests of modern astronomy. The
higher the redshift of the system, the younger it is, hence scientists
are continually searching for higher redshift galaxy
clusters. Moreover, from a very simplified point of view (probably too 
simple, however), the more massive the cluster, the longer it takes
to virialize and therefore when virialized, the older it is
(e.g. Sarazin 1986). A good way to detect clusters is to use X-rays (a
not exhaustive reference list is: Burke et al. 1997, Henry et al.
1997, Scharf et al. 1997, Ebeling et al.  1998, Rosati et al. 1998,
Vikhlinin et al. 1998, Nichol et al. 1999, Romer et al.  2000, Pierre 
et al. 2006). This
is because rich clusters should have a hot intra cluster medium (ICM
hereafter), which is detectable as a thermal X-ray source against the
relatively faint diffuse X-ray background.

X-ray searches for distant rich clusters have generally required that
the X-ray sources be detected as extended. In all cases, however, the X-ray
sources were followed up by optical imaging and spectroscopy (see
e.g. the Bright-SHARC survey: Romer et al. 2000) in order to obtain redshifts
and to characterize the galaxies and richness of the
clusters. Unfortunately, not all clusters appear as extended
X-ray sources if the point spread function of the X-ray telescope has
been degraded, which can be the case when images are far off
axis. Furthermore, the X-ray emission of distant clusters and groups
may have relatively small angular extents on the sky, and the emission
may be dominated by a central AGN or a cool bright X-ray core. The goal of the
project described in this paper was, therefore, to search for distant X-ray 
extended sources that do not
appear extended in Rosat PSPC data (due to instrument limitations), as used by
the SHARC surveys.

As distant ($\simgreat 0.8$) X-ray luminous clusters are rare, it is
necessary to search the largest possible region of the sky for which
deep X-ray exposures are available. We therefore used our previously 
reported Bright-SHARC survey (e.g. Romer et al. 2000). From the entire catalog
of sources detected in the ROSAT pointed observations (covering about
180 deg$^2$; the effective area decreases with the sensitivity level,
see Section~\ref{dis_con}), we removed:

- all the sources detected as extended in the PSPC data above 
$1.5\times 10^{-13}$ \ecmss\ found in the original Bright-SHARC survey (Romer
\etal, 2000); here our survey goes almost 10 times fainter and is therefore 
limited by definition to sources not appearing as extended; 

- all sources with a detection limit lower than $3.5\sigma$;

- all the sources identified with known single objects (single galaxies, QSOs
and stars) from the NED and Simbad databases;

- all the sources with a clear NVSS (Dickey $\&$ Lockman 1990) radio
source, to avoid as much as possible bright distant AGNs or
QSOs, within a radius of 2~arcmin. We also checked aposteriori
that no radio source from the Radio Master Catalog available at
http://heasarc.gsfc.nasa.gov/
was detected within our X-ray contours. More specifically, among the
38 entries of Table~1, 35 are covered by the FIRST radio survey (White
et al. 1997), but no radio source is located within our X-ray
contours.

- all the sources with a clearly visible optical object in the DSS
photographic plates (Lasker et al. 1990) within a radius of 2~arcmin.

We finally distilled down the original list of over 3,000 objects to 36 
candidates that we are in the
process of systematically examining with new deep optical imaging,
optical spectroscopy, and X-ray followup observations when possible.
We also verified there was no duplication between our source list and
what was available in the literature for these targets up to
Nov. 2006. These 36 remaining sources are then:

- either rich clusters, too distant and/or too faint and/or too far offaxis 
to be detected as
extended in our automated processing scheme. As seen in the images,
however, a few objects show possible evidence for extended
X-ray emission. These candidate clusters are also too distant to have their
galaxy population clearly visible in the DSS.  Assuming a magnitude of
${\rm M_R}=-23$ for a typical brightest cluster galaxy, the cosmological parameters
given at the end of this section and assuming a magnitude
limit of R$\sim$18 for the DSS, this gives a minimal redshift of 0.3
for these clusters (and lower is we assume they are groups instead of 
clusters, see the text for individual objects below);

- either distant X-ray active objects (e.g. AGNs, QSOs) or stars too
faint to be seen in the DSS.



Although the project is not complete, we have gathered a significant
amount of data that warrant a ``mid-term'' report.  It is interesting to
compare these results with newer ongoing surveys such as ChaMP
(Barkhouse et al. 2006) so that when designing future missions devoted
to large sky surveys it can be judged whether it is better to design a
telescope that cuts off at relatively low energies (similar to ROSAT)
versus a smaller field of view telescope but with significant
collecting area up to at least 7~keV (such as \cha\ or \xmm).

It is also important to note how followup ground based observations are
beginning to reveal sets of underlumnious X-ray clusters which have the
potential of complicating cosmological interpretations of S-Z and X-ray cluster
surveys.

We will assume for the purpose of calculations that H$_0$ = 71 km
s$^{-1}$ Mpc$^{-1}$, $\Omega _\Lambda =0.73$ and $\Omega _m=0.27$.


\section{The data}\label{sec:data}

We give the list of our 36 candidates in Table 1. This table
summarizes for each object the observational details described
hereafter.

\begin{table*}
\caption{36 X-ray sources (plus two additional structures) in the
survey. We give the coordinates (from the SHARC wavelet analysis, see
Adami et al. 2000), the optical imaging data origin and
characteristics (exposure time ET in minutes), the X-ray data (ROSAT
PSPC, XMM or Chandra and radial offset in arcmin), the optical spectroscopy data origin (S for SDSS, CA
for Calar Alto or V for the VIMOS IFU), the source status (Single:
X-ray point source, Cluster: galaxy structure, Pending: to be
determined), an estimated redshift (if a galaxy structure), the ROSAT PSPC
count rates (CR in units of 10$^{-3}$/s in the [0.5-2] keV range),
except for Cl~J1202+4439 where it is an XMM count rate in the [0.5-10] keV band; all the luminosities 
given are based on the assumption of the maximum estimated redshift, in units
of $10^{44}$~\ecmss, and ${\rm n_H}$ in units of $10^{22}$ cm$^{-2}$. The two additional sources
are not detected in the original SHARC analysis and we therefore do not give
X-ray properties for them.}
\begin{tabular}{ccccccccccc}
\hline
Source & $\alpha$(2000) & $\delta$(2000) & imaging & ET & X-rays & Sp &
Status & z & CR & L$_Xbol$/nH\\
\hline
Cl0223-0856 & 02 23 06.10 & -08 56 47.6 & OHP 1.2m R      & 60        & ROSAT 8& S           & Cluster? & 0.49?& 1.97 & 0.424/.029\\
Cl0240-0801 & 02 40 09.56 & -08 01 06.6 & SDSS $\chi ^2$        &              & ROSAT 19&                & Pending  & & 3.08 & \\
Cl0241-0802 & 02 41 03.56 & -08 02 11.3 & ESO 3.5m I      & 10      & ROSAT 12 &                & Cluster? & 0.55?& 2.71 & 0.762/.035\\
Cl0242-0756 & 02 42 22.17 & -07 56 04.9 & SDSS $\chi ^2$        &              & ROSAT 27&                & Single?  & & 7.98 & \\
Cl0254+0012 & 02 54 24.25 & +00 12 52.6 & DSSRed/Blue $\chi ^2$ &              & ROSAT 11& S          & Pending  & & 3.23 & \\
Cl0302-1526 & 03 02 41.77 & -15 26 47.0 & DSSRed/Blue $\chi ^2$ &              & ROSAT 17&                 & Single?  & & 2.25 & \\
Cl0317-0259 & 03 17 26.87 & -02 59 34.5 & DSSRed/Blue $\chi ^2$ & & ROSAT 8  &                & Single?  & & 1.89 & \\
 &  &  &  & & XCS &                &  & &  & \\
Cl0413+1215 & 04 13 54.03 & +12 15 58.8 & OHP 1.2m R        & 150       & ROSAT 13&                & Pending & & 2.43 & \\
Cl0922+6217 & 09 22 53.19 & +62 17 14.8 & OHP 1.2m R       & 40           & ROSAT 9&                & Pending  & & 2.73 & \\
Cl0937+6105 & 09 37 48.47 & +61 05 27.6 & ARC 3.5m i'      & 36       & ROSAT 20&                & Pending  & & 3.10 & \\
Cl~J1024+1935 & 10 24 23.89 & +19 35 15.8 & ARC 3.5m i'       & 50        & ROSAT 20&                & Cluster? & 0.15-0.65?& 6.09 &2.49/.0214\\
Cl~J1024+1943 & 10 24 37.92 & +19 43 14.9 & ARC 3.5m i'       & 90        & ROSAT 17&                & Pending  & & 3.10 & \\
Cl~J1050+6317 & 10 50 17.65 & +63 17 45.2 & ARC 3.5m i'      & 50       & ROSAT 25&                & Pending  & & 4.08 & \\
Cl~J1052+5655 & 10 52 11.89 & +56 55 35.3 & SDSS $\chi ^2$        &              & ROSAT 26& S           & Single? & & 10.69 & \\
Cl~J1052+5400 & 10 52 46.60 & +54 00 02.6 & ARC 3.5m i'      & 90      & XMM        12&                & Spurious? & & 2.02 & \\
Cl~J1102+2514 & 11 02 08.95 & +25 14 18.5 & ARC 3.5m i'       & 50        & ROSAT 12&                & Cluster? & 0.15-0.65?& 3.65 & 1.24/0.0140 \\
Cl~J1103+2458 & 11 03 21.80 & +24 58 49.8 & SDSS $\chi ^2$        &              & ROSAT 11&                & Pending  & & 1.31 & \\
Cl~J1113+4042 & 11 13 34.59 & +40 42 32.9 & ARC 3.5m i'       & 90        & Chandra    13& CA      & Cluster  & 0.51& 4.68 & 1.09/0.0184\\Cl~J1120+1254 & 11 20 48.59 & +12 54 58.8 & CFHT 3.6m B       & 20 & XMM     9  &                & Single   & & 3.53 & \\
            &  &  & CFHT 3.6m V       & 15 & XCS   &                &    & &  & \\
            &  &  & CFHT 3.6m R       & 10 &    &                &    & &  & \\
Cl~J1121+4309 & 11 21 40.67 & +43 09 06.8 & SDSS $\chi ^2$        &              & ROSAT 9&                & Pending  & & 3.14 & \\
Cl~J1121+0338 & 11 21 56.65 & +03 38 18.8 & ARC 3.5m r'    & 90     & ROSAT 30& S/V & Single   & & 15.18 & \\
            &  &  & ARC 3.5m i' band   & 80     & &  &    & &  & \\
Cl~J1158+5541 & 11 58 50.65 & +55 41 34.4 & SDSS $\chi ^2$        &              & XMM  12      &                & Pending  & & 1.92 & \\
Cl~J1202+4439 & 12 02 33.03 & +44 39 42.8 & SDSS $\chi ^2$        & & XMM 10 &                & Cluster?  & 0.28& 7.36 & 0.212/.0135 \\
 &  &  &         & & XCS &                &  & &  &  \\
Cl~J1207+4429 & 12 07 40.91 & +44 29 38.8 & SDSS $\chi ^2$        &              & ROSAT 12&                & Pending  & & 1.97 & \\
Cl~J1213+3908 & 12 13 32.88 & +39 08 24.7 & ARC 3.5m i'      & 30        & ROSAT 24&                & Single? & & 7.47 & \\
Cl~J1213+3317 & 12 13 53.75 & +33 17 27.4 & SDSS $\chi ^2$        &              & ROSAT 16&                & Pending  & & 2.46 & \\
Cl~J1214+1254 & 12 14 50.32 & +12 54 01.9 & ESO 3.5m I        & 10        & ROSAT 14&                & Pending & & 6.15 & \\
Cl~J1216+3318 & 12 16 22.89 & +33 18 28.5 & ARC 3.5m r'   & 90     & ROSAT 17&                & Pending  & & 7.82 & \\
Cl~J1216+3318 &  &  & ARC 3.5m i' band   & 90     & &                &  & &  & \\
Cl~J1234+3755 & 12 34 00.88 & +37 55 49.2 & SDSS $\chi ^2$        &              & ROSAT 20& S           & Single   & & 4.65 & \\
Cl~J1237+2800 & 12 37 18.90 & +28 00 16.5 & SDSS $\chi ^2$        &              & ROSAT 18&                & Pending & & 5.21 & \\
Cl~J1259+2547 & 12 59 20.71 & +25 47 10.4 & SDSS $\chi ^2$        &              & ROSAT 51&                & Pending  & & 3.15 & \\
Cl~J1343+2716 & 13 43 08.32 & +27 16 38.7 & SDSS $\chi ^2$        &              & ROSAT 13&                & Pending  & & 2.63 & \\
Cl~J1350+6028 & 13 50 45.95 & +60 28 39.2 & ARC 3.5m i'      & 50        & ROSAT 21&                & Single? & & 4.01 & \\
Cl~J1411+5933 & 14 11 08.37 & +59 33 12.5 & ARC 3.5m i'     & 50        & ROSAT 25&                & Cluster? & 0.25-1?& 5.30 & 5.97/0.0166\\
Cl~J1514+4351 & 15 14 11.33 & +43 51 23.9 & ARC 3.5m i'      & 20        & ROSAT 12& S           & Cluster? & 0.3-1?& 2.58 & 2.90\\
Cl~J1651+6107 & 16 51 02.95 & +61 07 25.3 & Gemini 8.2m r'     & 15     & XMM    9    & CA      & Cluster? & 0.2-0.5?& 2.72 & 0.608/.025 \\            &  &  & Gemini 8.2m i'     & 17      &  &       &  & &  &  \\
\hline
 & 10 50 30    & +63 19 18   & ARC 3.5m i' band   & 50        & ROSAT &     CA         & Cluster  & 0.53&            & \\    
 & 11 13 42    & +40 42 22   & ARC 3.5m i' band   & 90        & ROSAT &                & Cluster? & 0.15-0.6?&            & \\    
\hline
\end{tabular}
\label{tab:cand}
\end{table*}

\subsection{Optical imaging}

 We first observed deep $i'$ (and sometimes $r'$) images of 13
candidates at the ARC 3.5 meter telescope with SPIcam\footnote{see
http://www.apo.nmsu.edu for details}. Exposure times ranged from 20 to
90 minutes. Two other candidates were observed in the R band at the ESO
3.5 meter telescope with EFOSC2 with exposure times of 10 minutes. Three 
candidates were observed in the R band at the OHP 1.2 meter
telescope and CCD camera with exposure times between 40 and 150
minutes. We obtained B, V and R band observations for one candidate at
the CFHT with the CFH12K camera (B: 20 minutes, V: 15 minutes, R: 10
minutes). Finally, we observed one candidate with the Gemini north
telescope in $r'$ and $i'$ for 15 and 17 minutes, respectively.

For the 16 remaining candidates, 13 are covered by the SDSS and 3
by the DSS Red and Blue photographic plates (McLean et
al. 2000). Although the SDSS and DSS data are shallower than our
direct observations, by quadratically summing the data in all
available bands ($u$, $g'$, $r'$, $i'$, $z'$ for SDSS, and blue and
red for DSS), taking into account the typical noise in each band,
we then produced deeper images than those in the individual bands. This
method is not rigourously optimal to detect the faintest possible
objects (see e.g. Szalay et al. 1999 for a better but more complex
extraction method) and is not intended to product any calibrated magnitudes. 
We used this method because our goal was
to get a deep representation of the field object populations, in order
to decide which target should be followed with very deep images. The
up-to-date results of the optical imaging along with the superposed
X-ray contours are given in the on-line material. The images made by
superposing various bands will be refered to hereafter as ``$\chi^2$'' images.

\subsection{X-ray imaging}

The target selection was made on the basis of ROSAT PSPC images
treated by a automated procedure described in Romer et al. (2000). In
addition, we found in the \cha\ and \xmm\ archives X-ray data for 6
candidates (5 candidates with \xmm\ data and 1 with \cha\ data), but
often at the field edge, with low exposure times, or both. The images
in the Appendix were overlayed with XMM/Chandra data when available and
with ROSAT PSPC data if not. X-ray contours were computed for each
candidate at 1$\sigma$ intervals starting from the 3$\sigma$ level. These
levels were computed using the background estimated from the count rate 
in an arbitrarily empty region close to the X-ray source. The
ROSAT PSPC images were smoothed over a 1.5 arcmin Gaussian window prior 
to generating the contours.

 \subsection{Spectroscopy}

We checked the NED database for available redshifts in the considered
areas (at this writing). We also obtained single slit spectra at the
Calar Alto 3.5 meter telescope with the MOSCA
spectrograph\footnote{see
http://w3.caha.es/CAHA/Instruments/MOSCA/manual.html} for three
candidates (exposure times ranging between 30 minutes and 3
hours). Finally, we got a 12 hour VIMOS IFU (Le F\`evre \etal,
2003)\footnote{http://www.eso.org/instruments/vimos/} observation of
one candidate (but weather conditions were quite poor).

\section{Description of the 36 candidates}
\label{sec:list}

Here we discuss the possible nature of each candidate with the data we
have in hand. Our conclusions are summarized in Table~1 and the
corresponding figures are given in the on--line appendix.

When objects are visible in the optical images (excluding the $\chi^2$
images) within the X-ray contours, we used the brightest one to
compute a minimal redshift for the structure, assuming that this
object is the dominant galaxy of the structure. We assumed an absolute
magnitude of ${\rm M_R}=-23$ and ${\rm M_{i'}} = -23.5$ for the
dominant galaxy in each structure (these values are typical in nearby
clusters).\sethlcolor{red}

We also computed an X-ray flux in the [0.5-2] keV energy range using
WEBPIMMS and WEBSPEC, as well as L$_{Xbol}$  (using XSPEC) for the X-ray objects that
we considered as possible clusters. We assumed a mean kT of 3 keV (the
effect of the assumed kT over a reasonable temperature range of 2-10 keV is,
however, less than 10\%)
and a Mekal model, with a fixed metallicity Z = 0.3Z$_{\odot}$. We used the
hydrogen column density in the Galaxy for each relevant pointing (see Table 1).

\subsection{ARC imaging}


Thirteen candidates have been followed up with imaging at the ARC 3.5 meter
telescope at least in the $i'$ band.

{\bf Cl~J0937+6105} and {\bf Cl~J1050+6317}: The central X-ray
contours of Cl~J0937+6105 and Cl~J1050+6317 are not clearly associated
with any optical object.

{\bf Cl~J1024+1935} has a regular shape but we have no spectroscopy for this
candidate. The brightest optical object has a magnitude of $i'$=19.4 and is
possibly a quasar or an AGN similarly to the optical counterpart of
Cl~J1121+0338, but the X-ray countours can be typical of a nearby galaxy structure. 

{\bf Cl~J1024+1943} is an X-ray source associated with a complex optical object
population, that consists of both quite bright and very faint objects.

{\bf Cl~J1052+5400}: This field contains a rich galaxy
population. However, the ROSAT PSPC X-ray source was not detected in
our analysis of recent \xmm\ data. This source could be a variable
point source, as the image quality is not good enough to determine if
the source is extended. Since it is not present in the \xmm\ data, the
source could also be spurious.

{\bf Cl~J1102+2514} is a relatively strong X-ray source that is
well centered on a relatively bright non circular optical object
($i'$= 19.2) and the contours also include several
other optical counterparts. This X-ray source is probably a group or
a cluster between redshifts $z \sim 0.15$ and $\sim 0.65$.

{\bf Cl~J1113+4042} is a complex X-ray source also observed with \cha. The
X-ray data are not deep enough and too far off axis (13\arcmin, \cha, 
14\arcmin, ROSAT)  to allow spectral
investigations or to determine a statistically significant X-ray
extent. The area probably includes an AGN (the west source) and two
real galaxy structures (see Section \ref{add} for a discussion of the
east structure). We determined three redshifts close to the central X-ray
emission. Two of the main galaxies are probably associated with the
X-ray emission, are at a redshift of $\sim$0.5 and have an early type 
appearance. Cl~J1113+4042 is therefore probably a galaxy group at z$\sim$0.5. 

{\bf Cl~J1121+0338} is a relatively strong X-ray source (two times stronger than Cl~J1024+1935)
and is well centered on a QSO observed by the SDSS at z=0.839. We also obtained VIMOS
IFU spectroscopic data for this region. No galaxy concentration
appears in redshift space and we therefore conclude that the 
X-ray emission of Cl~J1121+0338 is due to a quasar at z$\sim$0.84.

{\bf Cl~J1213+3908} and {\bf Cl~J1350+6028} have outer X-ray contours
that include several peaks (the X-ray emission of Cl~J1213+3908 covers
the whole optical field of view) or are very large. The optical fields are dense enough,
however, that the coincidence between these inner contour peaks and
the optical objects could be purely by chance.

\begin{figure}
\centering
\includegraphics[width=8cm,angle=270]{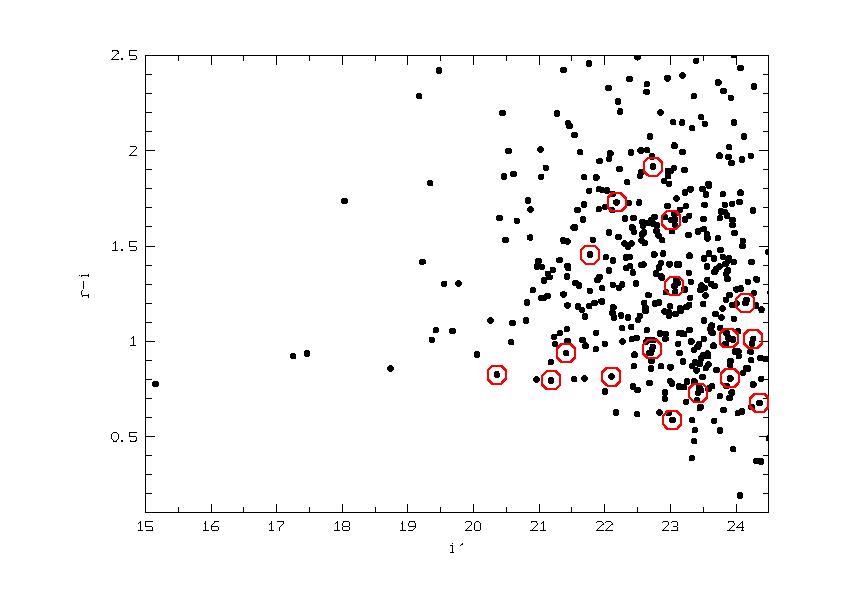}
\caption[]{$r'-i'$ vs. $i'$ color magnitude relation for the Cl~J1216+3318
field of view. The larger (red) filled circles are the galaxies
included within the X-ray contours.}
\label{fig:CMR36}
\end{figure}

{\bf Cl~J1216+3318} has a rich galaxy population that extends outside
the border of the X-ray contours.  The X-ray contours are quite
irregular and it is difficult to conclude with the data in hand
whether this a cluster or a single source. In order to explore further
the possible existence of a cluster we have plotted in
Fig.~\ref{fig:CMR36} the galaxy color magnitude relation in the
field. The large (red) circles are the optical objects within the
X-ray contours. When we also include objects outside the X-ray
contours, we find marginal evidence for a red sequence around
$r'-i'$=0.8 (11 objects among the 17 within the X-ray contours have
$0.5<r'-i'<1.1$), that would place a galaxy structure between z=0.2
and 0.5 (from Fukugita et al. 1995).  However, the absence of a
central bright galaxy makes it difficult to conclude if Cl~J1216+3318
is a cluster or single source.

{\bf Cl~J1411+5933} is an X-ray source with at least 5 optical objects within the
X-ray contours, the brightest one having an i' magnitude of 20.7. This
would place the galaxy structure at $z \sim$1 if it is an ${\rm M_{i'}}=-23.5$
galaxy associated to a rich cluster or at z$\sim$0.25 if it is an
${\rm M_{i'}}=-20$ group central galaxy. 

{\bf Cl~J1514+4351}: This X-ray source is quite elongated, hence it
is apparently an extended X-ray source. Several optical objects
(including at least two galaxies) are visible within the X-ray
contours. The brightest (and most extended one) has an $i'$ magnitude
of 20.9. This apparent magnitude places this cluster candidate at
z$\sim$1 if this is an M$_{i'}=-$23.5 galaxy associated with a rich
cluster, or at $z \sim$0.3 if this is an M$_{i'}=-$20 central galaxy
of a group. A foreground galaxy has been measured by the SDSS at
z=0.16518 and is associated with a larger foreground galaxy cluster at
this redshift. However, this foreground structure is probably not
related to the X-ray source. This is because the X-ray emission does not overlap
the $z \sim 0.16$ galaxy.

\subsection{Gemini data for Cl~J1651+6107}

This candidate has a relatively low Galactic latitude
($\sim37^{\circ}$) and is located in a region populated by Galactic
stars and with a prominent galactic H$\alpha$ emission coming from the
Draco cloud (e.g. Penprase et al. 2000).  This is confirmed by the two
spectra of cold stars (located at the upper left and lower right in
Fig.~36 of the online data) we obtained at Calar Alto (with MOSCA)
within the Cl~J1651+6107 field of view. The two bright objects embedded in the
diffuse optical emission (see Fig.~36 in the Appendix) are also stars,
as deduced from the $r'$ Gemini image.

\begin{figure}
\centering
\includegraphics[width=8cm,angle=270]{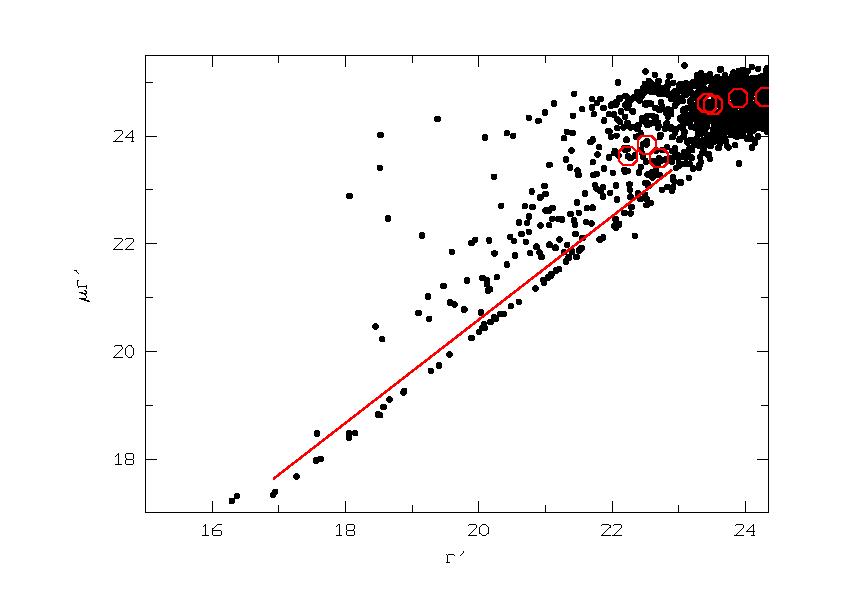}
\caption[]{Cl~J1651+6107: central surface brightness versus total $r'$ magnitude 
diagram used to distinguish stars from galaxies, see text. }
\label{fig:sg17}
\end{figure}

\begin{figure}
\centering
\includegraphics[width=8cm,angle=270]{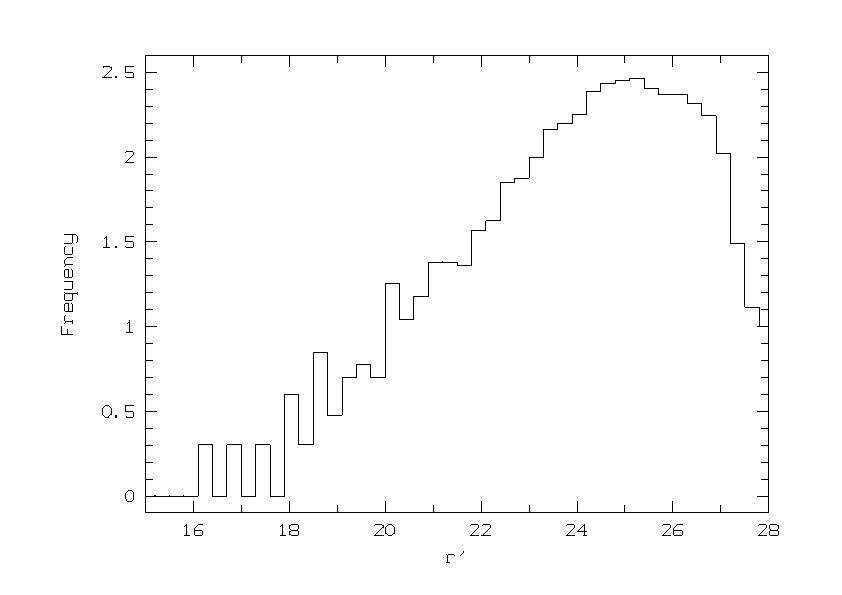}
\caption[]{LOG-normal Cl~J1651+6107 $r'$ magnitude histogram for the Gemini field,
suggesting completeness up to $r'\sim$24.5.}
\label{fig:histo17}
\end{figure}

\begin{figure}
\centering \includegraphics[width=8cm,angle=270]{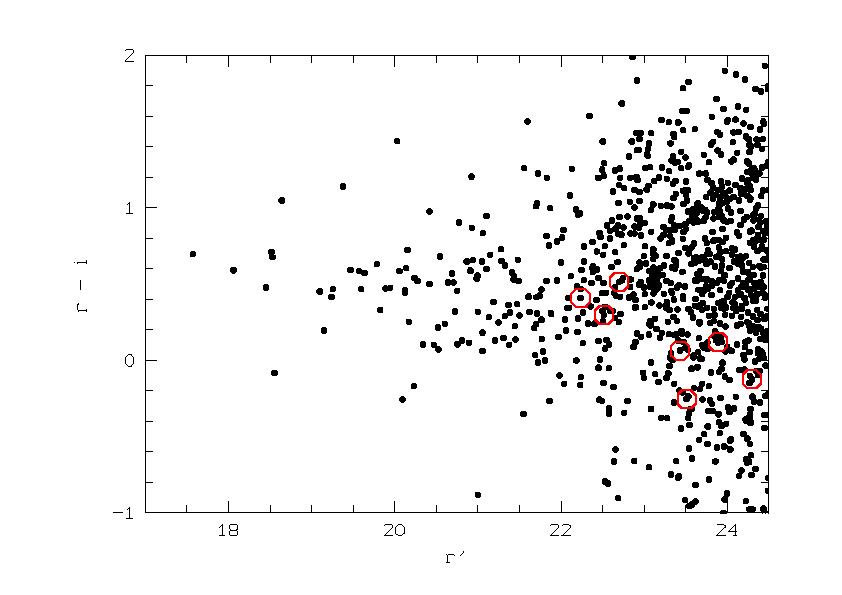}
\caption[]{Cl~J1651+6107: $r'-i'$ color versus $r'$ magnitude for objects
classified as galaxies. The 7 large (red) circles correspond to the
galaxies within the X-ray contours.}
\label{fig:cmr17}
\end{figure}

We found XMM data in the archive for this candidate, but the exposure
time is far too low to allow any spectral analysis. The X-ray emission
is, however, located on the top of 7 very faint objects that are galaxies, 
based on the imaging data used in Figs.~\ref{fig:sg17}, \ref{fig:histo17}
and \ref{fig:cmr17}. 

Fig.~\ref{fig:sg17} shows a clear star--galaxy separation down to $r' \sim
22.9$. We considered all objects
fainter than $r' \sim$ 22.9 as galaxies, since the Galactic star
contribution at these magnitudes is very low (e.g. Adami et al.
2006a).

These results suggest that we have found a structure of
galaxies.  In order to estimate its redshift, we limited our analysis
of the $r'$ data to $r'=24.5$ (as suggested by Fig.~\ref{fig:histo17})
and we plotted in Fig.~\ref{fig:cmr17} the color magnitude relation of
all objects classified as galaxies.  The three brightest galaxies
within the X-ray contours have $r'-i'$ colors close to 0.4. Following
Fukugita et al. (1995) and assuming these are early type galaxies,
this would place the structure between z=0.2 and 0.5. The fainter
objects are probably very low mass objects which were not able to
retain most of their metals and appear therefore quite blue
(e.g. Adami et al. 2006b).

Given the magnitude of the brightest galaxy within the X-ray contours
one interpretation is that this structure is a group, since this
magnitude is too faint to be a cluster dominant galaxy at
z$\leq$0.5. The absolute $r'$ magnitude of the brightest galaxy would
be $-17.7$ at z=0.2 and $-20.0$ at z=0.5. The absolute magnitudes are 
in the range of L$^*$ values for groups or clusters.
An alternative explanation is that Cl~J1651+6107 is a
cluster at $z \sim 1$ if our interpretation of the colors did not
produce the true value of the redshift.

\subsection{CFHT data for Cl~J1120+1254}

\begin{figure}
\centering
\includegraphics[width=8cm,angle=0]{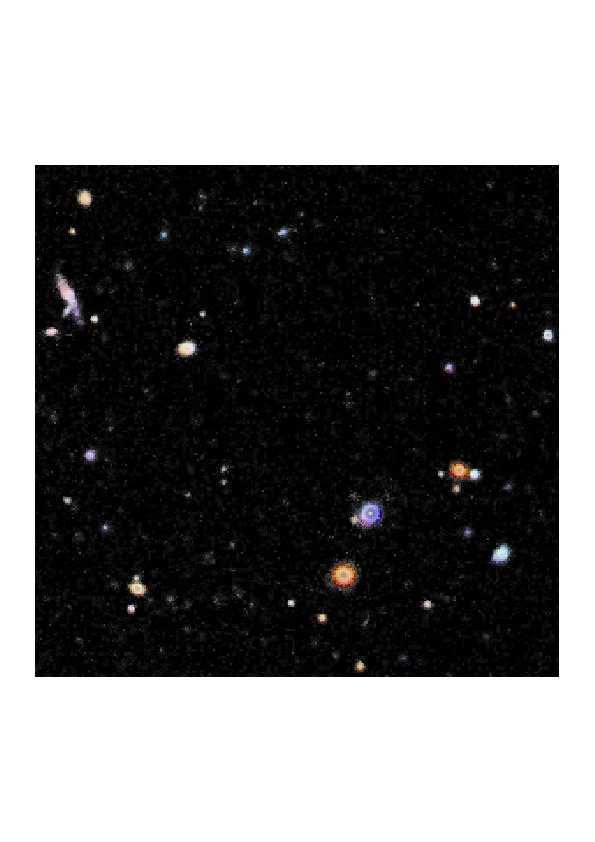}
\caption[]{BVR color image of Cl~J1120+1254. The two bright objects at the
image bottom (one blue, one red) are the brightest objects within the
X-ray contours.}
\label{fig:col13}
\end{figure}

We have obtained B, V and R CFHT CFH12K images for {\bf Cl~J1120+1254}.  The
images show several objects inside the X-ray contours. One of the two
brightest galaxies has a very blue color, the other a very red color
(Fig.~\ref{fig:col13}).

\begin{figure}
\centering
\includegraphics[width=8cm,angle=270]{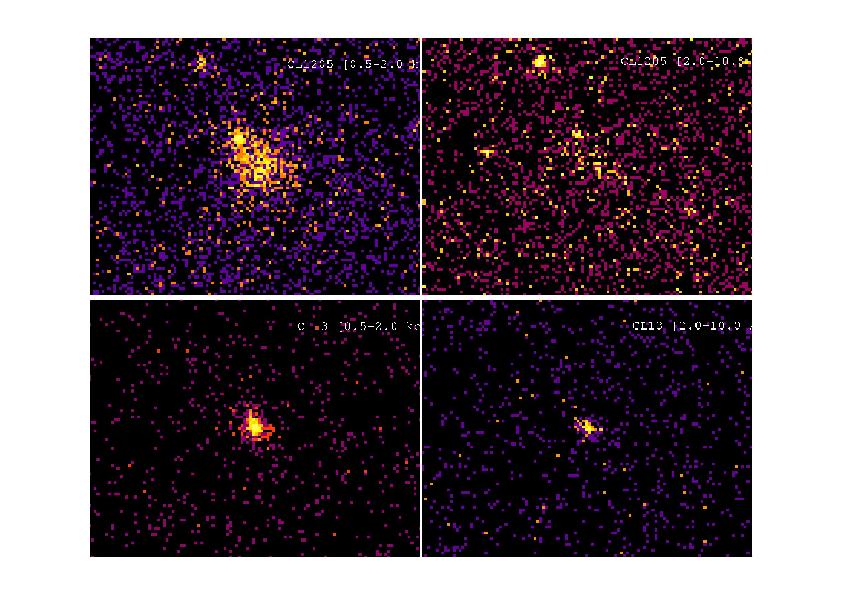}
\caption[]{ A known galaxy structure, ClG J1205+4429 (top) with a 
line-of-sight X-ray detectable AGN in the upper left of the image, see
Ulmer et al. 2005. [0.5-2.0] keV (left) and [2.0-10.0] keV band (right). 
Similar images of Cl~J1120+1254 (bottom).}
\label{fig:hs13}
\end{figure}

There are XMM data available for this candidate but the source is
located off axis ($\sim 9'$), so the angular resolution is degraded by
about 50\% (this XMM source is however detected as not extended by the ongoing
XCS survey, Romer et al., private communication and 2001). For
display purposes, we made two images, one in the [0.5-2.0] keV band
(soft) and the other in the [2.0-10.0] keV band (hard). Clusters are
expected to appear stronger in the soft band than in the hard band,
while AGN should look point-like in both bands. Following this, we have 
plotted in Fig.~\ref{fig:hs13} both
the Cl~J1120+1254 X-ray source and the X-ray image of a known galaxy structure 
(ClG J1205+4429,
hereafter Cl~J1205, see Ulmer et al. 2005). The image of this 
known galaxy structure
also has a very prominent AGN in its field (at the north east in the
image). Cl~J1205 (the known galaxy structure) is a strong source in the soft
band and a very weak one in the hard band, while the AGN can be seen to be 
relatively strong in both bands (although weaker in the soft band than in 
the hard band). In contrast, Cl~J1120+1254 also appears to be relatively
strong in both bands (even if weaker in the hard band compared to
the soft band). Given its X-ray image, however, we conclude that Cl~J1120+1254 
is probably an X-ray point source, despite that the spectral shape is not the
same as that of the AGN in the field of Cl~J1205.

\subsection{ESO 3.5m EFOSC2 data for Cl~J0241-0802 and Cl~J1214+1254}

{\bf Cl~J0241-0802} and {\bf Cl~J1214+1254} have been imaged with the
ESO 3.5 meter telescope and the EFOSC2 instrument (imaging mode). Both
candidates are associated with relatively bright optical objects.

{\bf Cl~J0241-0802} is probably a galaxy structure given the large number of
optical sources and the ``dominant galaxy'' appearance of the
brightest object visible within the X-ray contours. This object has an
I magnitude of 19, placing the possible galaxy structure at
z$\sim$0.55. The X-ray contours also suggest that this is an
extended X-ray source. At such a redshift, its extent corresponds to a
diameter of 500~kpc, in good agreement with the extent of a typical
group of galaxies (or poor cluster).

{\bf Cl~J1214+1254} is poorer than Cl~J0241-0802 from an optical point of view but
its X-ray shape seems extended. It is however impossible to conclude
on the nature of this source with the data in hand.

\subsection{OHP data for Cl~J0223-0856, Cl~J0413+1215 and Cl~J0922+6217}

These three cluster candidates were observed at the OHP 1.2 meter
telescope.

The ROSAT X-ray images of Cl~J0223-0856 and Cl~J0413+1215 are relatively round. 

\begin{figure}
\centering
\includegraphics[width=8cm,angle=270]{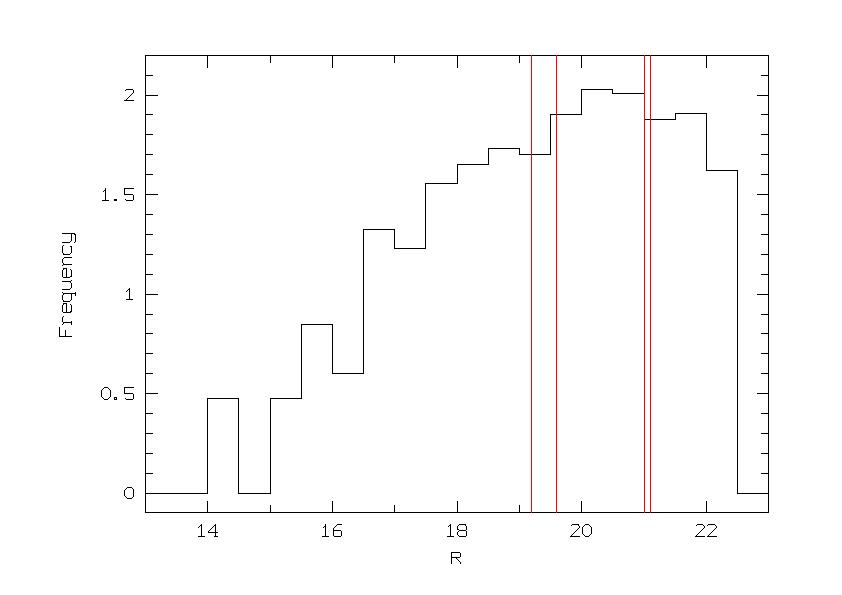}
\caption[]{LOG-normal Cl~J0223-0856 R magnitude histogram. The
  vertical (red) lines are the four objects detected within the X-ray
  contours.}
\label{fig:histo01}
\end{figure}

In the R image of {\bf Cl~J0223-0856} there are four faint objects visible within
the X-ray contours (see image in online data Fig.~1 and magnitude
histogram in Fig.~\ref{fig:histo01}).  If Cl~J0223-0856 is a cluster
of galaxies, then its redshift could be z=0.49 based on the brightest
detected galaxy. The angular extent of the X-ray emission is PSF
dominated but is equivalent to a 200 kpc diameter circle at this
redshift. This is compatible with its being a cluster core or a
group. Cl0223-0856 could therefore be a galaxy structure.

There is only one optical object visible within the X-ray contours of
{\bf Cl~J0413+1215}, but the source could be a cluster of galaxies as the
magnitude limit of the R band image is about 21 which means it is
unlikely that fainter cluster member galaxies would be visible in the
R image.  If Cl~J0413+1215 is indeed a cluster of galaxies, its redshift could be
z$\sim$1 based only on the single detected galaxy (R = 21.7) within
the X-ray contours. The size of the X-ray emission is also PSF
dominated and is equivalent to a 300 kpc diameter circle at this
redshift. The data are, however, too sparse to allow us to make a definitive
statement about its nature.

{\bf Cl~J0922+6217} has faint optical objects within its contours. The X-ray source
could be either a cluster of galaxies or an X-ray point source.

\subsection{SDSS data}


Thirteen of our cluster candidates have no deep CCD imaging. For each of
these objects, we summed in quadrature all the SDSS available bands
($u, g', r', i', z'$) to make a visible band image onto which to
overlay the X-ray contours. 

{\bf Cl~J0240-0801, Cl~J1103+2458, Cl~J1207+4429} and {\bf
Cl~J1343+2716} all have faint optical objects within their
contours. These X-ray sources could be either clusters of galaxies or
X-ray point sources.

{\bf Cl~J1052+5655} is possibly made up of a
collection of X-ray point sources. Within this region one galaxy has a
measured redshift of z=0.52147 (taken from NED), but it falls at the border of
the X-ray contours. The data are too noisy to determine if
the X-ray source is truly extended or not. It is probably a
collection of individual sources.

{\bf Cl~J1121+4309} has a quite regular and PSF dominated X-ray shape. 
There is no visible object within ROSAT PSPC X-ray contours. The optical 
counterpart is very faint.

{\bf Cl~J1158+5541} has XMM data, but the image is located at the edge of the MOS
fields ($\sim$12' off axis) and is not in the PN field. This prevents
us from deriving an X-ray spectrum.  The PSF is badly degraded at
this location and we cannot provide a reliable measure of the extent
of this X-ray source either. This candidate is however associated with a
faint optical object population, and thus its nature is indeterminate between
a distant cluster and an AGN.

\begin{figure}
\centering
\includegraphics[width=8cm,angle=270]{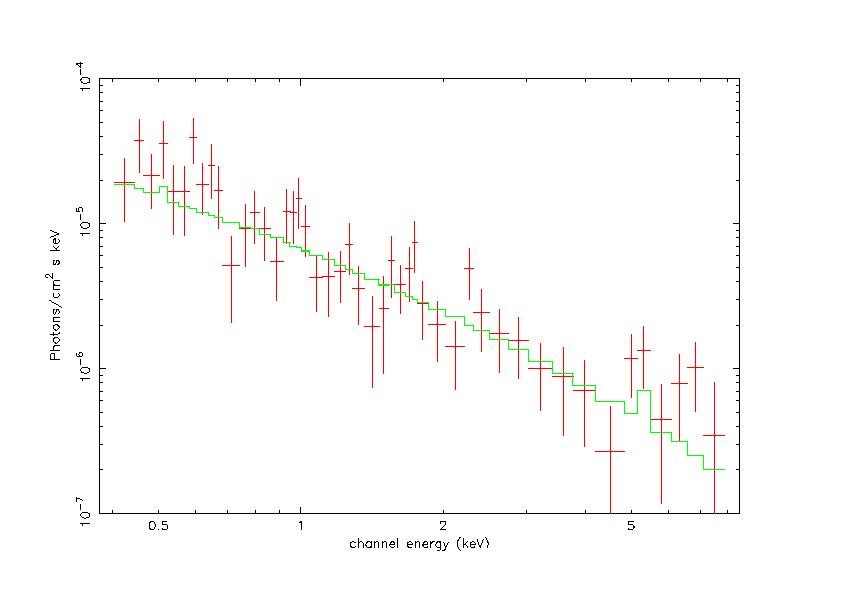}
\caption[]{X-ray photon spectrum of Cl~J1202+4439.}
\label{fig:spec28}
\end{figure}

{\bf Cl~J1202+4439} is a relatively strong X-ray source (S/N greater than 6 in the
ROSAT PSPC data) for which \xmm\ data (net exposure time of 36.7 ksec
after flare removal) are also available. This source was not detected as
extended by the ongoing XCS survey (Romer et al.: private communication and 
2001).
Our \xmm\ analysis generated
$\sim$300 photons in the source after background removal;
we produced an X-ray spectrum (Fig.~\ref{fig:spec28})
and fit a MEKAL model (N$_H$=1.35 10$^{20}$ cm$^{-2}$ and fixed
metallicity of 0.3 Z$_\odot$). The best fit was obtained for
a redshift of
0.28$^{+0.21}_{-0.28}$. The spectrum in Fig.~\ref{fig:spec28} shows a
$\sim$1$\sigma$ feature that could be due to Fe emission at about 5.2 keV, but
many other similarly sized features exist in the spectrum.  The 5.2 keV energy
for the rest frame 6.7 keV Fe line is consistent with the derived redshift. 
There are two optical objects within the X-ray contours which have magnitudes
consistent with being galaxies at this $\sim 0.3$ redshift. The
object is then possibly a group of galaxies at z$\sim$0.3 with an
estimated luminosity of 1.1 10$^{43}$ erg/s (in the [0.5-10] keV
range) but the nature of the structure is still to be confirmed with optical 
spectroscopy. This would be typical of a bright galaxy group (e.g.  Jones et
al. 2003).

{\bf Cl~J1213+3317} and {\bf Cl~J1237+2800} consist of several large X-ray sources with embedded faint
optical objects. These candidates could be either clusters of galaxies
or made of several unrelated X-ray point sources.

{\bf Cl~J1234+3755} is possibly an extended X-ray source. There is an SDSS QSO on the edge of
the X-ray contours (at z=0.57313), so some of the X-ray emission could
originate from this QSO.  We cannot exclude the
possibility of having a QSO embedded in a cluster.

{\bf Cl~J0242-0756} and {\bf Cl~J1259+2547} are made up of weak X-ray sources with a few faint
optical objects within their contours. Cl~J1259+2547 seems more extended than Cl~J0242-0756.

\subsection{DSS2 red and blue data for Cl~J0254+0012, Cl~J0302-1526 and Cl~J0317-0259}

These 3 candidates have no CCD imaging at all. We only used the
quadratically summed DSS2 Red and Blue photographic plate data to
overlay the ROSAT PSPC X-ray contours. 

{\bf Cl~J0254+0012} appears to be a collection of X-ray point sources.
There is one galaxy at the edge of the field with an SDSS
redshift ($z$=0.35952) and the cluster of galaxies SDSS CE
J043.601063+00.230312 has been detected at z=0.32 (estimated by
the SDSS teams as indicated in NED) also at the edge of the field.
We note that we used DSS2 data for Cl~J0254+0012 and not SDSS data
because this source is only located about 1 arcmin south of an SDSS
covered area.

{\bf Cl~J0302-1526} appears to be a collection of X-ray point sources with one
apparent optical identification and is probably not a cluster of
galaxies.

{\bf Cl~J0317-0259} has an X-ray emission that is PSF dominated with one visible optical
object within the X-ray contours. This source has also been detected in the
ongoing XCS survey (Romer et al.: private communication and 2001) as an 
unextended source.
This source is therefore possibly a real point source (given the better XMM
angular resolution compared to the ROSAT PSPC) but its nature remains undeterminate.

\subsection{Additional candidates}
\label{add}

\begin{figure}
\centering
\includegraphics[width=8cm,angle=0]{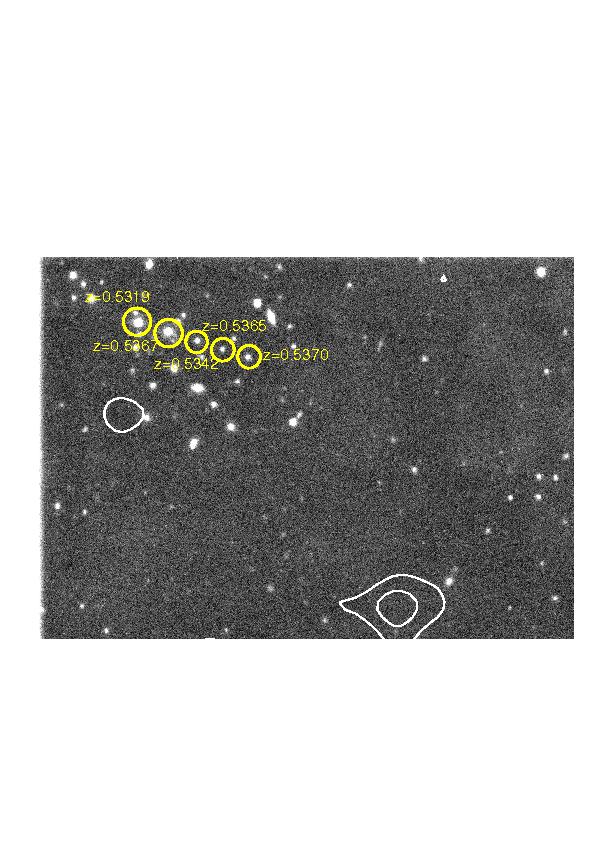}
\caption[]{A serendipitously detected galaxy structure in the Cl~J1050+6317 field of
view. White contours are X-ray ROSAT PSPC data.} 
\label{fig:Cl1050+6317bis}
\end{figure}

We found by chance another cluster of galaxies in the {\bf
Cl~J1050+6317} field of view (Fig.~\ref{fig:Cl1050+6317bis}). This
structure is clearly visible in the optical but completely invisible
in the ROSAT PSPC data, implying that it is under luminous in X-rays.
We measured five redshifts with the MOSCA instrument at the Calar Alto
3.5 meter telescope. All proved to be around z=0.535 and they were
mainly characterized by absorption lines (only one clearly shows the
[OII] emission line). Five redshifts are not enough to give a robust
velocity dispersion, but the raw computation gives a value of 430
km/s. This structure is probably a moderately massive cluster. The
brightest galaxy has an $i'$ magnitude of 20.1, corresponding to an
absolute magnitude of $-22.3$ which is typical for the central galaxy
of a relatively rich galaxy structure.

We also probably found another galaxy structure east of the {\bf
Cl~J1113+4042} ROSAT X-ray source. This X-ray structure has not been
sampled with optical spectroscopy but clearly appears associated with
a galaxy concentration of a dozen galaxies. The brightest galaxy of
this X-ray source has an $i'$ magnitude of 19.3 (yielding a redshift
in the [0.15-0.6] redshift range, depending on whether it is
a group or a rich cluster).

\section{Discussion and Conclusions} 
\label{dis_con}  


We have given an update on our extended SHARC survey. Due to the large
number of sources and significant amount of observing time needed to
obtain redshifts and new X-ray measurements of all the objects, we
have presented this work as an intermediate report, so as to make the
data available to the public. For simplicity we will assume in our
discussion that all the galaxy concentrations we have found are rich
clusters, but we acknowledge that some of these could be groups or
concentrations on the line of sight of active galaxies and are not necessarily
gravitationally bound systems.

There are several interesting aspects to this work: (a) optical 
cluster searches versus X-ray observations, (b) how we
compare with the recent ChaMP results (Barkhouse et al. 2006, also a
work in progress); (c) how this relates to future missions designed to
find clusters and/or AGN; and (d) the QSO population we found.

Both Donahue et al. (2006) and Barkhouse et al. (2006) demonstrated
that it is possible to find optical or near IR concentrations of
galaxies that are probably clusters of galaxies, but that these can be
weak X-ray emitters (see also Stanford et al. 2005). In our work, we
have found the same, in that simply taking relatively deep (i.e. with
exposure times $\sim 90$ minutes in the $i'$-band with 4-m class
telescopes) $3\arcmin \times 3\arcmin$ images can reveal clusters of
galaxies in the $z = 0.5 -0.7$ range (see e.g. the serendipitously discovered
cluster in the field of Cl~J1050+6317). Our $i'$ images were also
large enough to encompass an area well outside the X-ray image
location, and we uncovered faint X-ray clusters and point sources in 
this process.


As shown by Brodwin \etal\ (2006), by moving further into the IR even more
distant clusters can be found and photometric redshifts can be estimated.
These low X-ray luminosity clusters may pose a potential puzzle: if they
are massive, then their baryon fraction must be small compared to
low redshift clusters. If this is the case, why is the hot gas missing?
Hence, a possible quandary arises for those who want to use either X-ray or
S-Z surveys to determine cluster evolution and for those using clusters as
cosmological probes. For if these clusters have indeed significant amounts
of matter, then these X-ray and S-Z invisible clusters must be taken into
account when comparing predictions of cluster evolution with cosmological
models. Thus picking out a set of these under-luminous $z \sim 0.5$--0.6
clusters from currently available data bases and measuring their velocity
dispersions and/or gravitational lensing signal to determine masses will be
very important when using cluster surveys to determine cosmological model
parameters. This proposed project would create a census of the X-ray under luminous
clusters to determine how their numbers compare to those of X-ray luminous
clusters.

\begin{figure}
\centering
\includegraphics[width=8cm,angle=0]{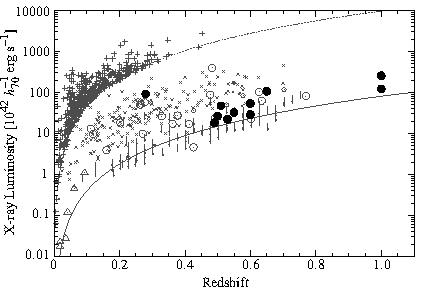}
\caption[]{Figure 10 of Barkhouse et al. (2006) showing the
distribution of X-ray luminosity (0.5-2.0 keV) as a function of
redshift. Open circles are the Barkhouse et al. (2006) extended X-ray sources
associated with clusters and the open triangles are extended X-ray sources
associated with nearby galaxies. Other symbols are from literature studies
(see Barkhouse et al. (2006) for details). Our data are shown with large 
black filled circles.}
\label{fig:Bark}
\end{figure}

Besides the optical and near IR observations, X-ray observations have been one
of the standard methods used for finding clusters of galaxies. The examination
of the field outside the pointing center of X-ray observations has also
been used for a long time (e.g. Henry et al. 1992). There are too many references
to review and compare with all the results of these works. We
therefore confine ourselves to comparing our work with a very recent survey by
Barkhouse et al. (2006), who surveyed 13 deg$^2$ down to a flux limit of about
$1.5 \times 10^{-14}$ \ecmss. They found $\sim$2.5 X-ray cluster candidates per deg$^2$ with
no optical counterpart. We estimate the areal coverage of our current 
survey to be $\sim$15 deg$^2$ from Fig.~12 of Adami et al. (2000) at the sensitivity 
level of about $2.3 \times 10^{-14}$ \ecmss. Thus, we have found comparable 
numbers of X-ray cluster candidates compared to the Barkhouse et
al. survey: 1.8 per deg$^2$. This value is based on a flux limit of 
$2.3 \times 10^{-14}$ \ecmss\ (compared to the $1.5 \times 10^{-14}$ \ecmss\ 
of Barkhouse et al. 2006), and on all the confirmed clusters
plus the candidates with a pending status in Table 1. 
Furthermore, in Fig.~\ref{fig:Bark}, it can also be seen that the redshift
distribution of the clusters and cluster candidates of our work and of 
Barkhouse et al. are similar.

This naturally leads to the question for future surveys as to what is
the best approach in terms of overall design for an X-ray telescope.
For example, the proposed VADER mission concept (Fassbender et al.  2006)  
uses the flight spare \xmm\ mirrors that would
allow an expanded and curved detector array to cover an extended field
of about 1 sq. degree. Although the \xmm\ detectors cover a 1~deg$^2$
FOV, the vignetting caused by the outer mirror graze angle (inner
mirrors have even smaller graze angles) of about 30~arcmin results in
a region of about $20\times 20$~arcmin$^2$ where the effective area is
$\simgreat 50$\%.  In contrast, a telescope that only works below
about 2 \keV\ such as ROSAT, but with an expanded area made possible
by using thin ($\sim$0.5 mm) electro formed mirrors, for example,
would be able to cover nearly $1 \times 1$ deg$^2$ with minimal
vignetting and a focal length about 2-2.5 times shorter (important
both for cost and for reducing the detector background for extended
sources); with a design to optimize the off axis angular resolution,
the survey would be approximately 10 times the coverage as the VADER
design (e.g.  Harvey et al. 2004; Atanassova \& Harvey 2003; 
Citterio \etal, 1999; Ulmer 1995; and Burrows et al. 1992).

It is the distant clusters ($ z \simgreat 1$) that are most important
to find, and their apparent temperature, $T_{\rm apparent} = 
T_{\rm intrinsic}/(1 + z)$ will be approximately 3 keV (or less), 
corresponding to intrinsic temperatures of 6 keV (or less). 
Hence most of these distant clusters can be found without
using X-ray telescopes that have significant effective area above
about 6 \keV. We have also shown that even with reduced off axis
angular resolution, clusters can be found.  The major drawback to a
mission with new mirrors is having to design in detail and then
fabricate new mandrels and mirrors.  In comparison, for the VADER
mission, the X-ray telescopes already exist, and the full array of IR
to Optical to X-ray telescopes combine to make the proposed
capabilities of the VADER mission concept impressive.

AGN/QSOs surveys would also benefit from an X-ray survey mission even
if it is one that only works below about 2 \keV. Although
AGN/QSOs may be heavily absorbed (e.g. have spectra that decrease
with decreasing energy due to photoelectric absorption in the rest
frame at about 4~\keV), QSOs (at least optically identified ones) show
a peak in their redshift distribution at about $z = 2$ (e.g. Schneider
\etal, 2005). The 4~keV absorption will be moved down to
1.3 \keV\ for these QSOs.  Also, many of those at even lower redshifts
should be easily found with a 0.5-2 \keV\ survey.


In conclusion, a combination of deep $i'$ band images and X-ray images  is a
productive way to find more clusters.  Some of the $i'$-band images (or a deep
$i'$ survey)  will likely produce X-ray faint (and under luminous) clusters that
are not coincident with the X-ray image. The under luminous X-ray clusters could
give us new understanding of both the formation of the Intra Cluster Medium (or the lack there
of) and the number of massive systems which could be missed by using X-ray (or
S-Z) surveys alone.  Furthermore, future missions that are aimed at searching
for  both distant rich clusters and QSOs in X-rays should seriously consider a
wide field of view (1$\times$1 deg$^2$ or more) telescope design optimized for
the $\sim 0.5-2$ \keV\ range.

\begin{acknowledgements} 

The authors thank the referee for his/her comments.
We thank J.C. Cuillandre for providing us with the CFHT/CH12K data for
Cl~J1120+1254 and M.A. Hosmer for comparisons with the XCS-DR1 
database. We also thank Calar Alto Observatory for allocation of
director's discretionary time to this programme.  This paper is based
on observations: 1)~Obtained with the Apache Point Observatory 3.5 m
telescope, owned and operated by the Astrophysical Research
Consortium. 2)~Obtained at the Canada-France-Hawaii Telescope (CFHT)
operated by the National Research Council of Canada, the Institut
National des Sciences de l'Univers of the Centre National de la
Recherche Scientifique of France, and the University of Hawaii.
3)~Obtained at the Gemini Observatory, which is operated by the
Association of Universities for Research in Astronomy, Inc., under a
cooperative agreement with the NSF on behalf of the Gemini
partnership: the National Science Foundation (United States), the
Particle Physics and Astronomy Research Council (United Kingdom), the
National Research Council (Canada), CONICYT (Chile), the Australian
Research Council (Australia), CNPq (Brazil) and CONICET (Argentina).
4)~Observations collected at the German-Spanish Astronomical Center,
Calar Alto, jointly operated by the Max-Planck-Institut f\"ur
Astronomie Heidelberg and the Instituto de Astrof\'\i sica de
Andaluc\'\i a (CSIC).  5)~Observations made with ESO Telescopes at the
La Silla and Paranal Observatories.  6)~SDSS data:
Funding for the SDSS and SDSS-II has been provided by the Alfred
P. Sloan Foundation, the Participating Institutions, the National
Science Foundation, the U.S. Department of Energy, the National
Aeronautics and Space Administration, the Japanese Monbukagakusho, the
Max Planck Society, and the Higher Education Funding Council for
England. The SDSS Web Site is http://www.sdss.org/. The SDSS is
managed by the Astrophysical Research Consortium for the Participating
Institutions. The Participating Institutions are the American Museum
of Natural History, Astrophysical Institute Potsdam, University of
Basel, University of Cambridge, Case Western Reserve University,
University of Chicago, Drexel University, Fermilab, the Institute for
Advanced Study, the Japan Participation Group, Johns Hopkins
University, the Joint Institute for Nuclear Astrophysics, the Kavli
Institute for Particle Astrophysics and Cosmology, the Korean
Scientist Group, the Chinese Academy of Sciences (LAMOST), Los Alamos
National Laboratory, the Max-Planck-Institute for Astronomy (MPIA),
the Max-Planck-Institute for Astrophysics (MPA), New Mexico State
University, Ohio State University, University of Pittsburgh,
University of Portsmouth, Princeton University, the United States
Naval Observatory, and the University of Washington.  Also based on
observations made at Observatoire de Haute Provence (CNRS), France.

This research has made use of the SIMBAD database, operated at CDS,
Strasbourg, France, and of the NASA/IPAC Extragalactic Database (NED)
which is operated by the Jet Propulsion Laboratory, California
Institute of Technology, under contract with the National Aeronautics
and Space Administration.

\end{acknowledgements}

\Online

\begin{appendix} 

We present in this appendix the 36 optical images of our candidates 
overlayed with X-ray contours (ROSAT data except when quoted). We also 
give the known redshifts in the given optical area.

\begin{figure}
\centering
\includegraphics[width=8cm,angle=0]{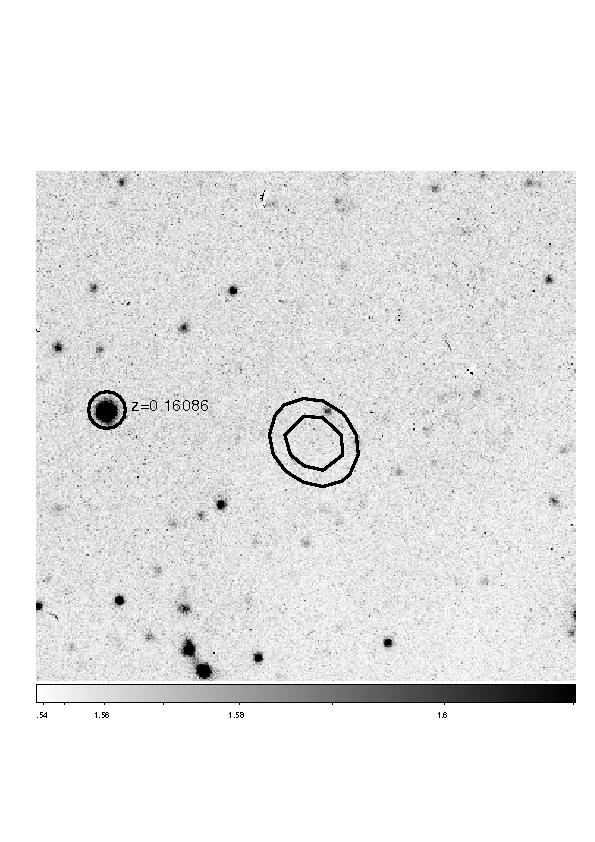}
\caption[]{R band OHP image for Cl~J0223-0856 (completeness level: R$\sim$20).
The field is $4.2\times4.2$ arcmin$^2$.}
\label{fig:cl01}
\end{figure}

\begin{figure}
\centering
\includegraphics[width=8cm,angle=0]{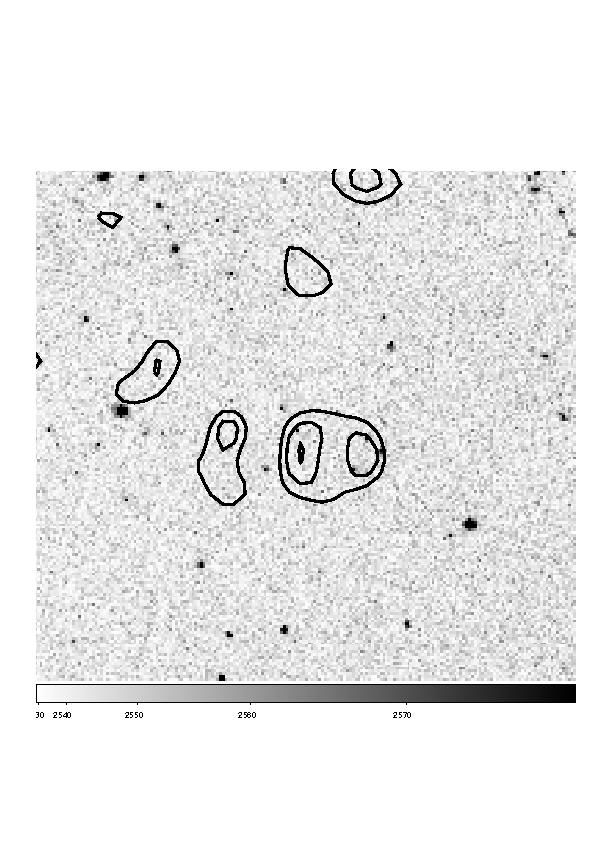}
\caption[]{$\chi^2$ image for Cl~J0240-0801 built from the SDSS u, g', r', i' and z' 
images. The field is 3.7$\times$3.7 arcmin$^2$.}
\label{fig:cl09}
\end{figure}

\begin{figure}
\centering
\includegraphics[width=8cm,angle=0]{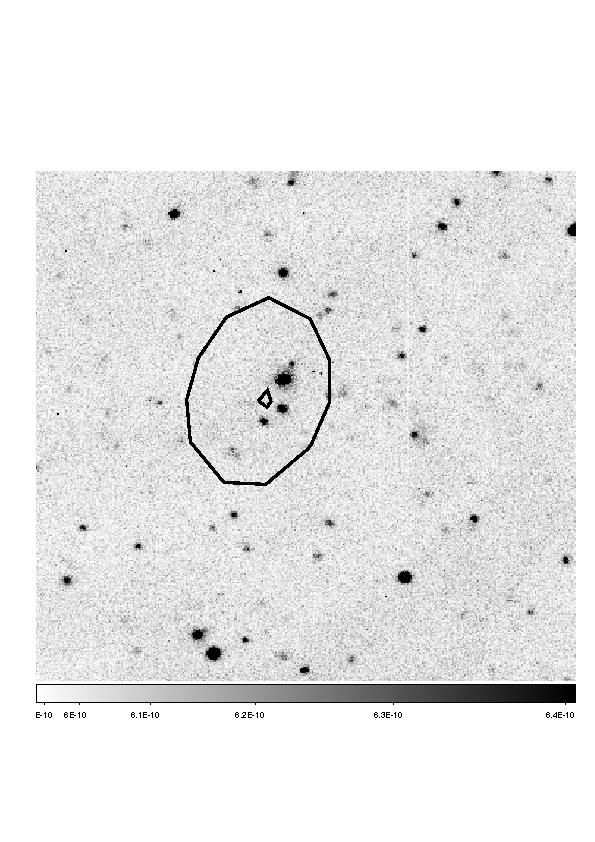}
\caption[]{I image for Cl~J0241-0802 observed at ESO (completeness level:
I$\sim$21). The field is 2$\times$2 arcmin$^2$.}
\label{fig:cl08}
\end{figure}

\begin{figure}
\centering
\includegraphics[width=8cm,angle=0]{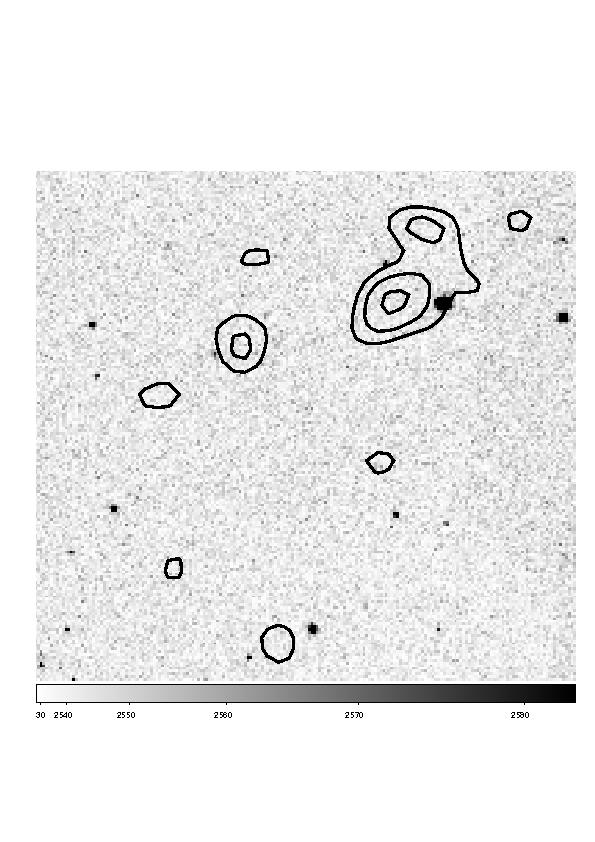}
\caption[]{$\chi^2$ image for Cl~J0242-0756 built from the SDSS u, g', r', i' and z' 
images. The field is 3.7$\times$3.7 arcmin$^2$. }
\label{fig:cl25}
\end{figure}

\clearpage

\begin{figure}
\centering
\includegraphics[width=8cm,angle=0]{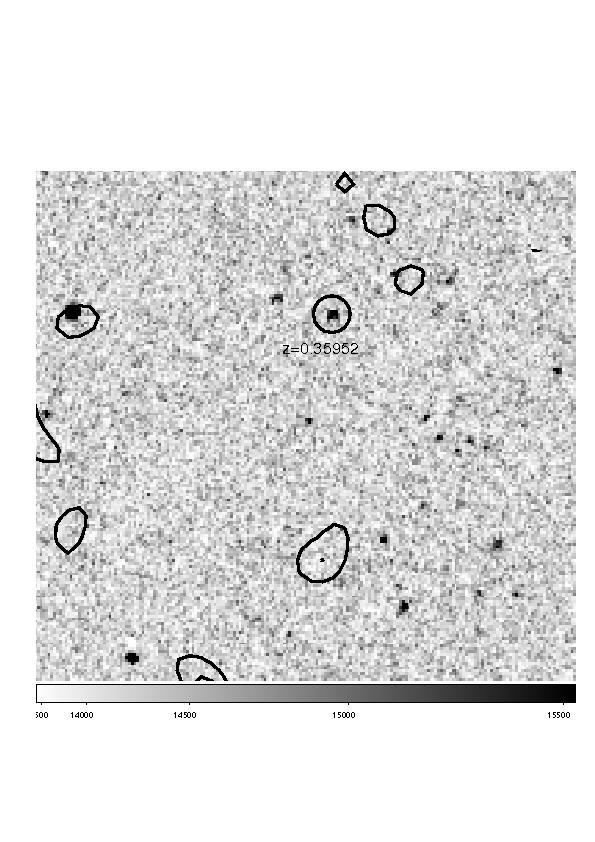}
\caption[]{$\chi^2$ image for Cl~J0254+0012 built from the DSS Red and Blue images. 
The field is 3.7$\times$3.7 arcmin$^2$.}
\label{fig:cl10}
\end{figure}

\begin{figure}
\centering
\includegraphics[width=8cm,angle=0]{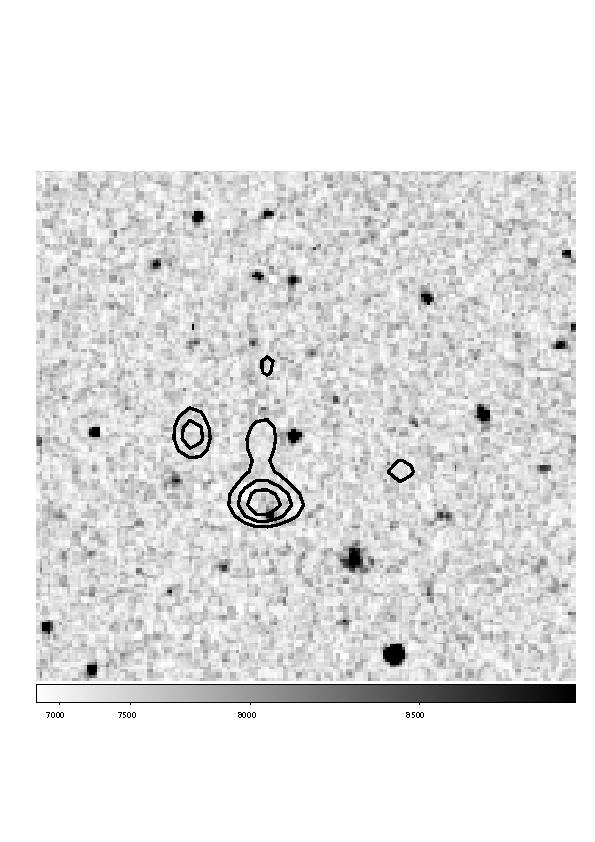}
\caption[]{$\chi^2$ image for Cl~J0302-1526 built from the DSS Red and Blue images. 
The field is 3.7$\times$3.7 arcmin$^2$.}
\label{fig:cl03}
\end{figure}

\begin{figure}
\centering
\includegraphics[width=8cm,angle=0]{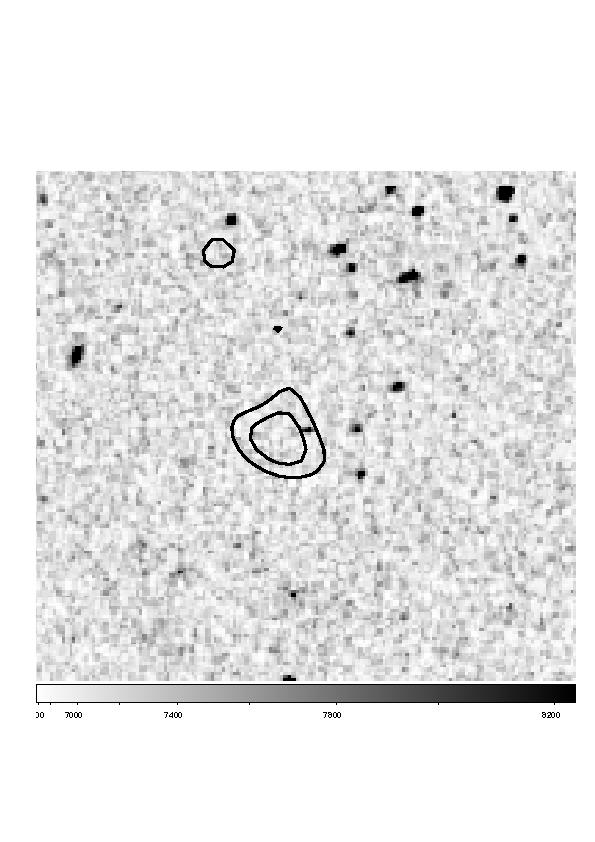}
\caption[]{$\chi^2$ image for Cl~J0317-0259 built from the DSS Red and Blue images. 
The field is 3.7$\times$3.7 arcmin$^2$.}
\label{fig:cl02}
\end{figure}

\begin{figure}
\centering
\includegraphics[width=8cm,angle=0]{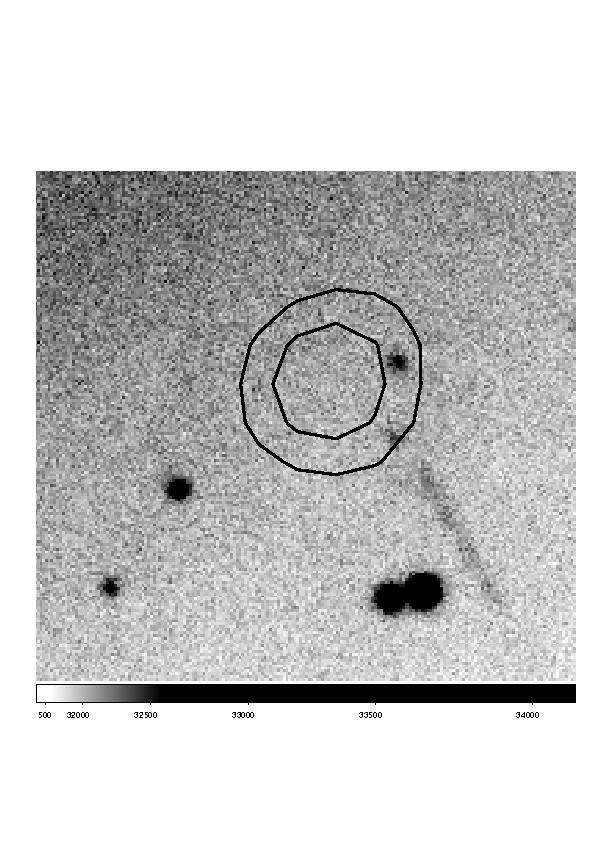}
\caption[]{R band OHP image for Cl~J0413+1215 (completeness level:
R$\sim$21.5). The field is 4.2$\times$4.2 arcmin$^2$.}
\label{fig:cl01}
\end{figure}

\clearpage

\begin{figure}
\centering
\includegraphics[width=8cm,angle=0]{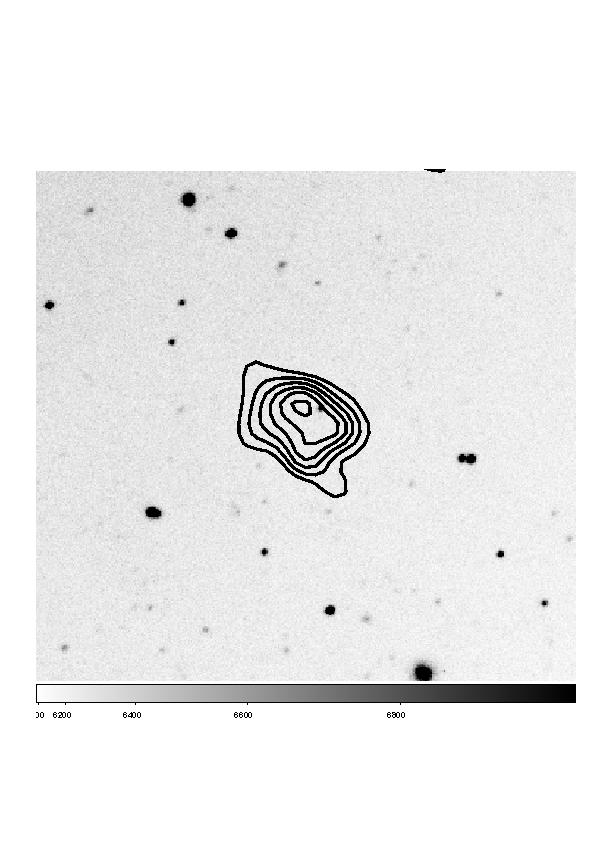}
\caption[]{R band OHP image for Cl~J0922+6217 (completeness level:
R$\sim$20). The field is 4.2$\times$4.2 arcmin$^2$.}
\label{fig:cl19}
\end{figure}

\begin{figure}
\centering
\includegraphics[width=8cm,angle=0]{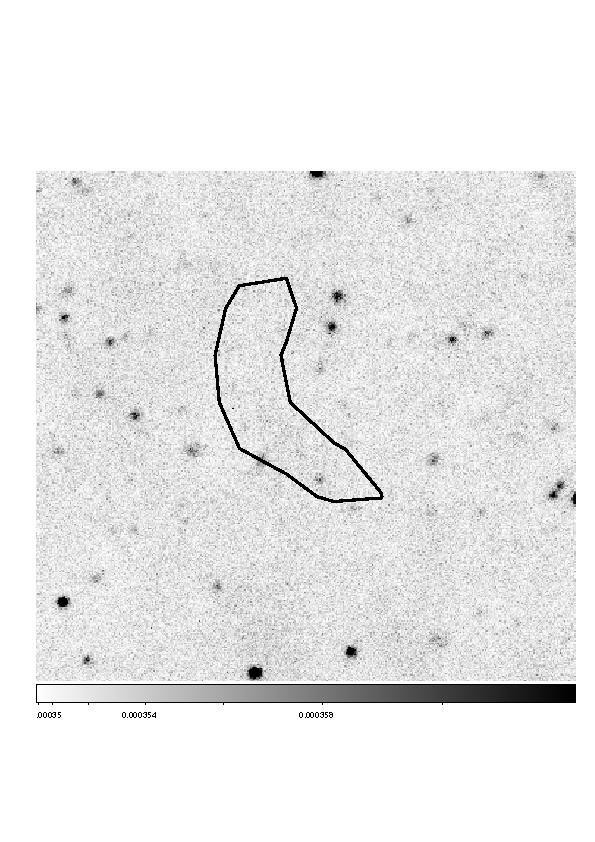}
\caption[]{i' image for Cl~J0937+6105 observed at ARC (completeness level:
 i'$\sim$22.5).  The field is 1.8$\times$1.8 arcmin$^2$.}
\label{fig:cl11}
\end{figure}

\begin{figure}
\centering
\includegraphics[width=8cm,angle=0]{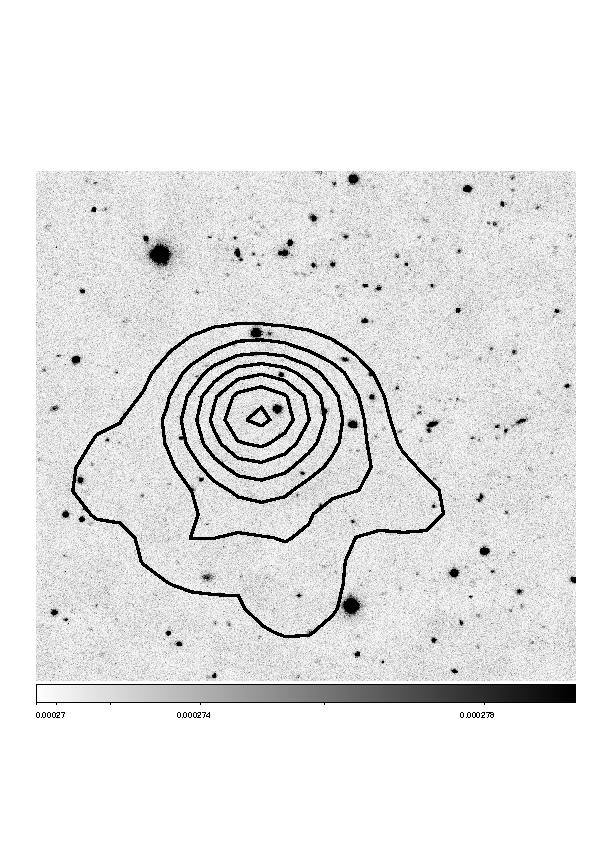}
\caption[]{i' image for Cl~J1024+1935 observed at ARC (completeness level:
i'$\sim$23).  The field is 3$\times$3 arcmin$^2$. }
\label{fig:cl30}
\end{figure}

\begin{figure}
\centering
\includegraphics[width=8cm,angle=0]{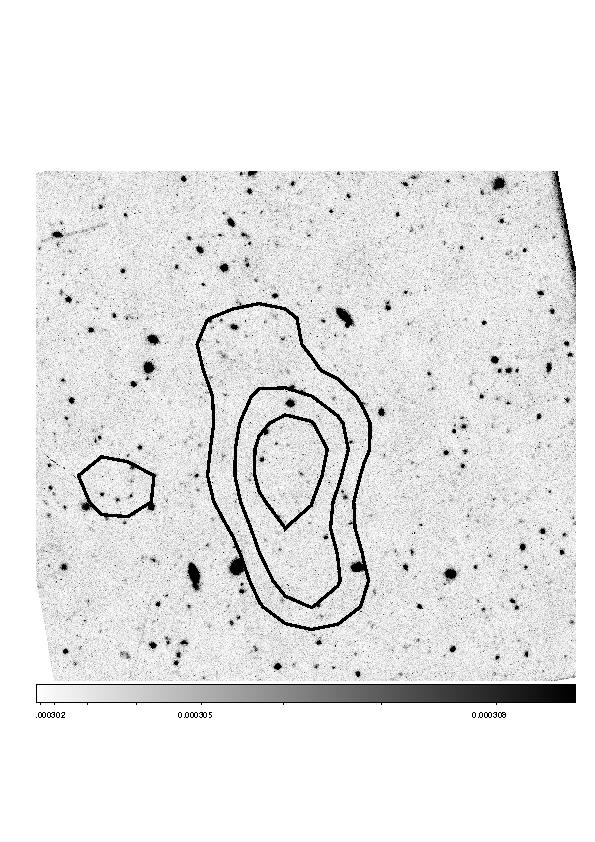}
\caption[]{i' image for Cl~J1024+1943 observed at ARC (completeness level:
  i'$\sim$23.5).  The field is 3$\times$3 arcmin$^2$.}
\label{fig:cl20}
\end{figure}

\clearpage

\begin{figure}
\centering
\includegraphics[width=8cm,angle=0]{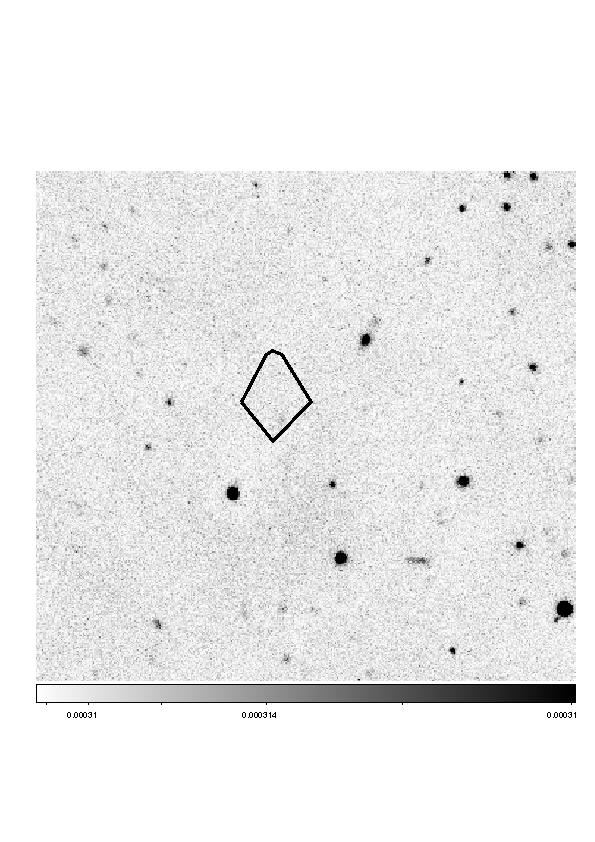}
\caption[]{i' image for Cl~J1050+6317 observed at ARC (completeness level:
 i'$\sim$23).  The field is 1.8$\times$1.8 arcmin$^2$.}
\label{fig:cl12}
\end{figure}

\begin{figure}
\centering
\includegraphics[width=8cm,angle=0]{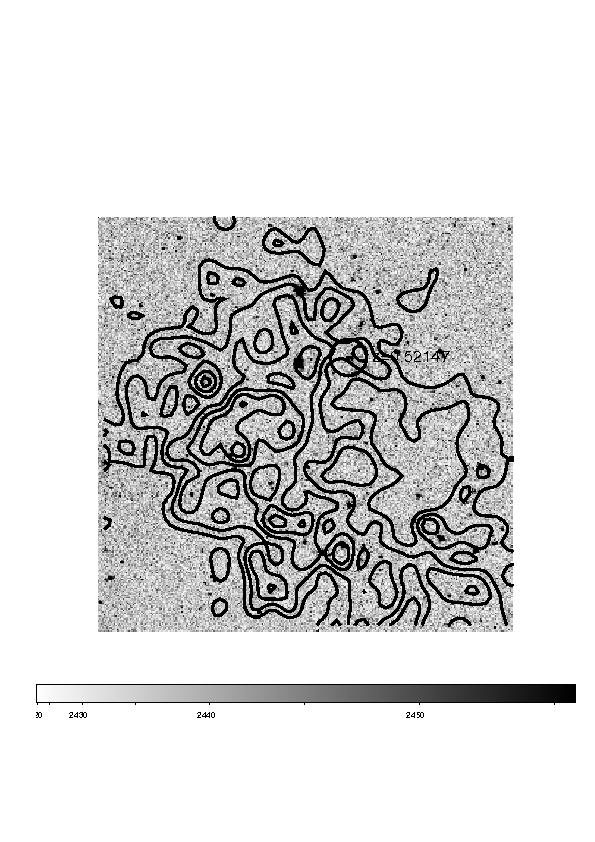}
\caption[]{$\chi^2$ image for Cl~J1052+5655 built from the SDSS u, g', r', i'
and z' images. The field is 5.9$\times$5.9 arcmin$^2$. }
\label{fig:cl33}
\end{figure}

\begin{figure}
\centering
\includegraphics[width=8cm,angle=0]{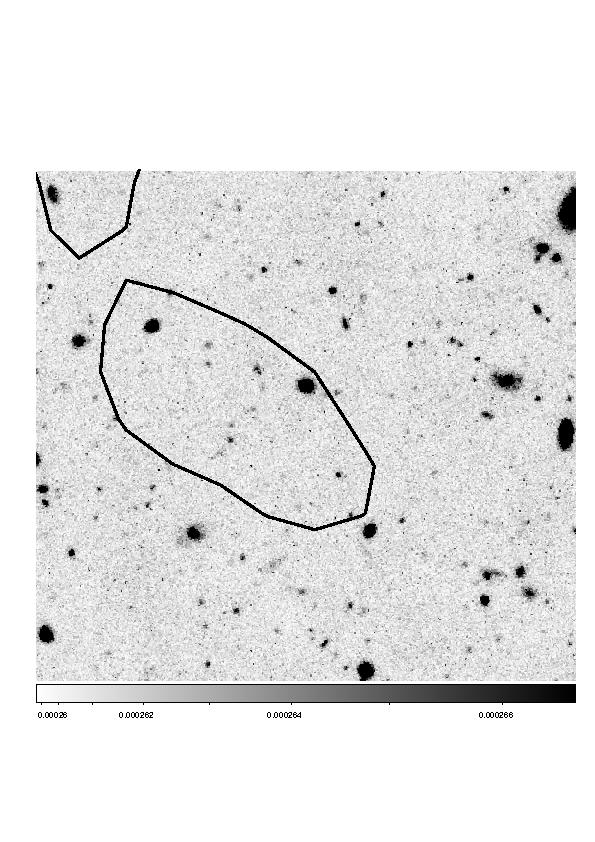}
\caption[]{$i'$ image for Cl~J1052+5400 observed at ARC (completeness level:
i'$\sim$24).  The field is 1.8$\times$1.8 arcmin$^2$.  The X-ray source
is from ROSAT data but is not confirmed by the available XMM data.}
\label{fig:cl04}
\end{figure}

\begin{figure}
\centering
\includegraphics[width=8cm,angle=0]{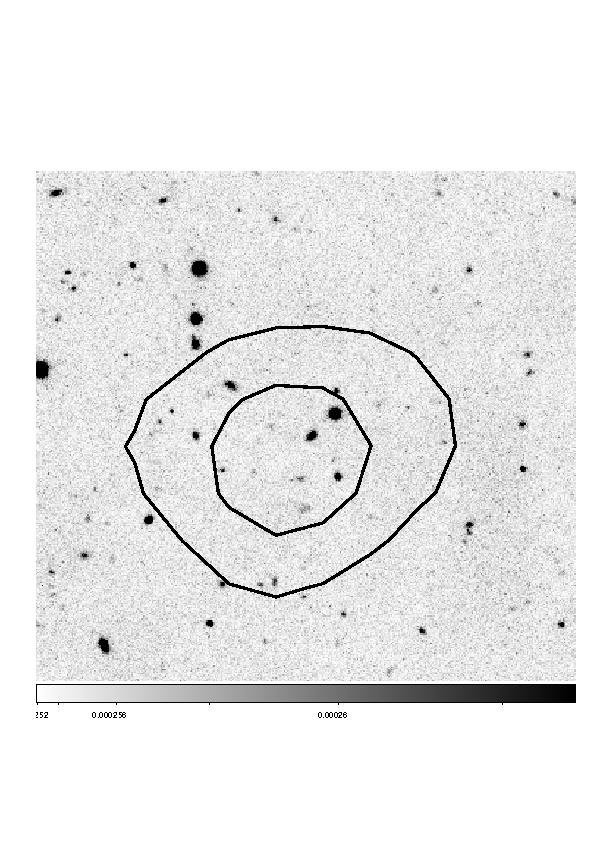}
\caption[]{i' image for Cl~J1102+2514 observed at ARC (completeness level:
i'$\sim$23).  The field is 1.8$\times$1.8 arcmin$^2$. }
\label{fig:cl35}
\end{figure}

\clearpage

\begin{figure}
\centering
\includegraphics[width=8cm,angle=0]{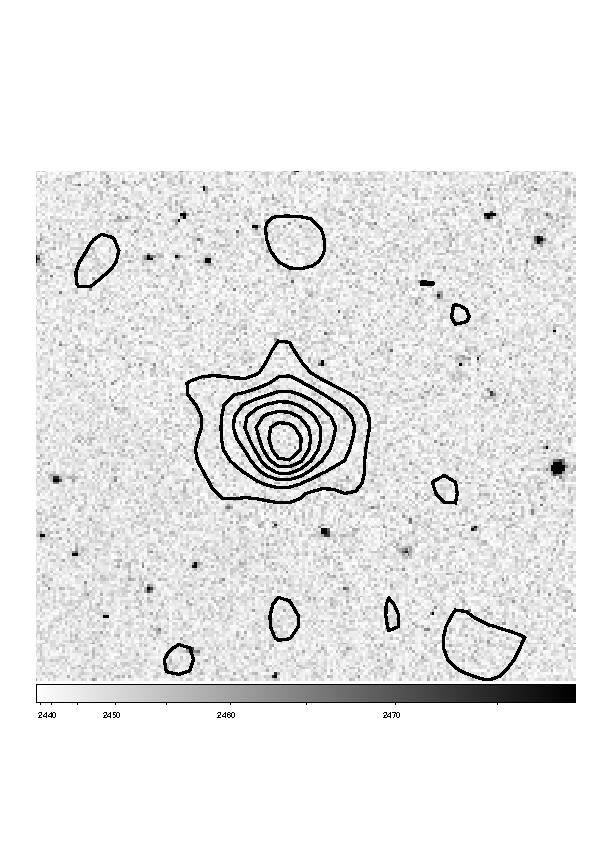}
\caption[]{$\chi^2$ image for Cl~J1103+2458 built from the SDSS u, g', r', i' and z' 
images. The field is 3.7$\times$3.7 arcmin$^2$. }
\label{fig:cl22}
\end{figure}

\begin{figure}
\centering
\includegraphics[width=8cm,angle=0]{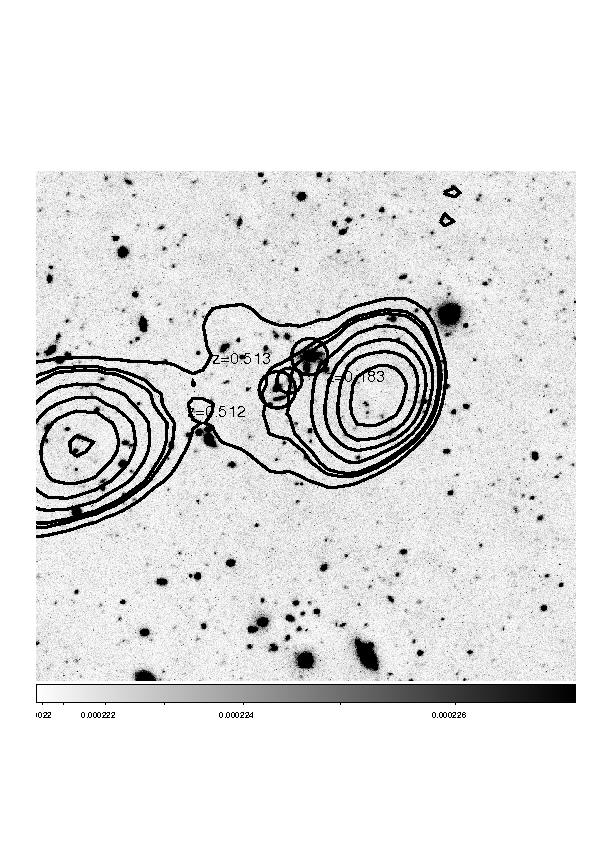}
\caption[]{i' image for Cl~J1113+4042 observed at ARC (completeness level:
i'$\sim$23.5).  The field is 3$\times$3 arcmin$^2$. Overlayed X-ray
contours are Chandra data.}
\label{fig:cl26}
\end{figure}

\begin{figure}
\centering
\includegraphics[width=8cm,angle=0]{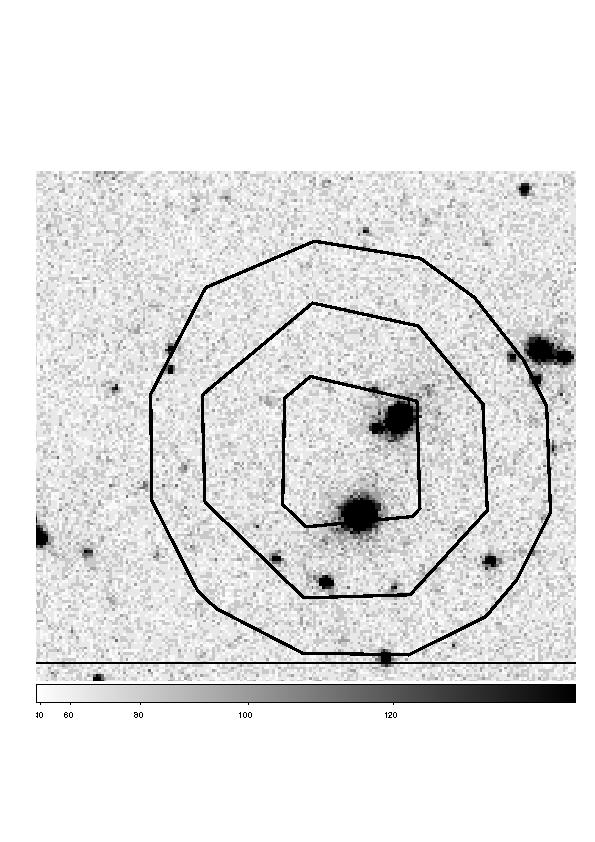}
\caption[]{R image for Cl~J1120+1254 observed at CFHT (completeness level:
R$\sim$22).  The field is 1.8$\times$1.8 arcmin$^2$. Overlayed X-ray
contours are XMM data.}
\label{fig:cl13}
\end{figure}

\begin{figure}
\centering
\includegraphics[width=8cm,angle=0]{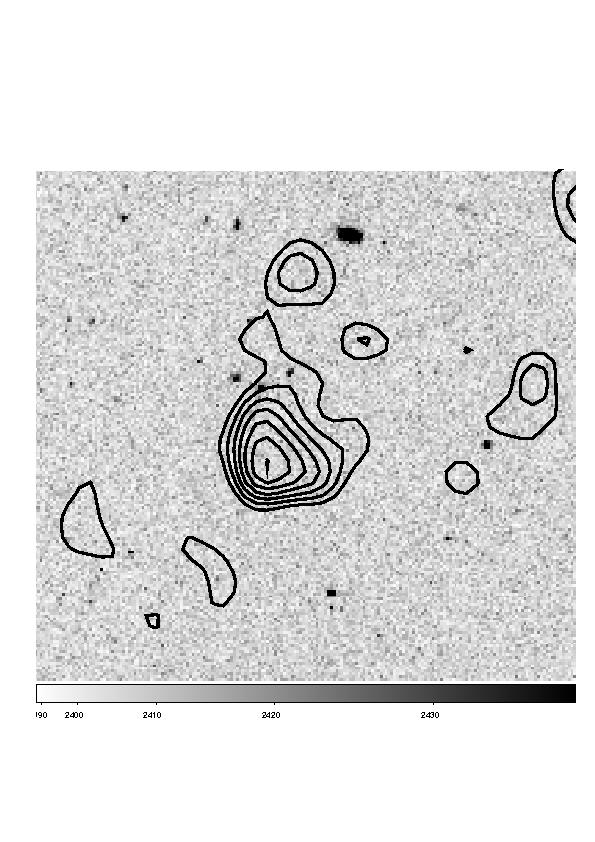}
\caption[]{$\chi^2$ image for Cl~J1121+4309 built from the SDSS u, g', r', i' and z' 
images. The field is 3.7$\times$3.7 arcmin$^2$.}
\label{fig:cl05}
\end{figure}

\clearpage

\begin{figure}
\centering
\includegraphics[width=8cm,angle=0]{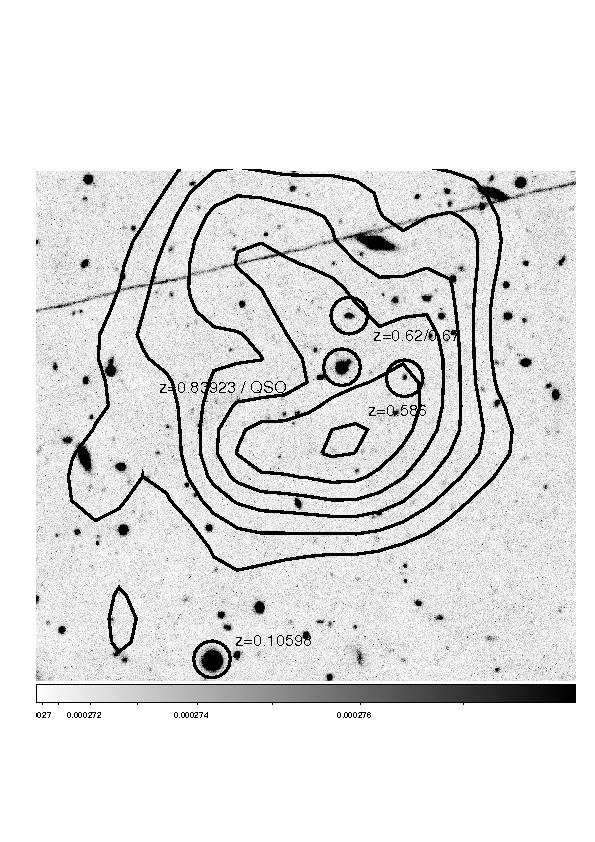}
\caption[]{i' image for Cl~J1121+0338 observed at ARC (completeness level:
i'$\sim$23.5).  The field is 3$\times$3 arcmin$^2$. }
\label{fig:cl34}
\end{figure}

\begin{figure}
\centering
\includegraphics[width=8cm,angle=0]{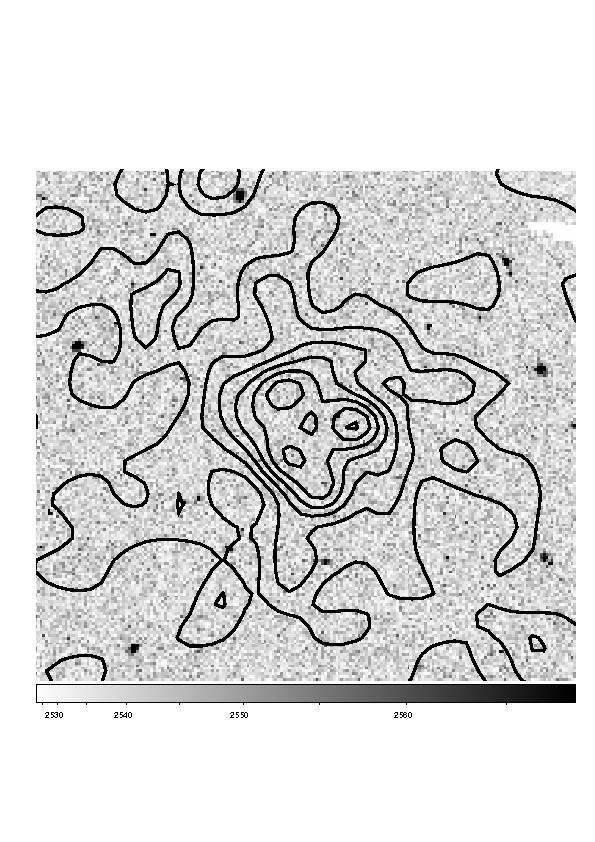}
\caption[]{$\chi^2$ image for Cl~J1158+5541 built from the SDSS u, g', r', i'
and z' images. The field is 3.7$\times$3.7 arcmin$^2$. Overlayed X-ray
contours are XMM data.}
\label{fig:cl21}
\end{figure}

\begin{figure}
\centering
\includegraphics[width=8cm,angle=0]{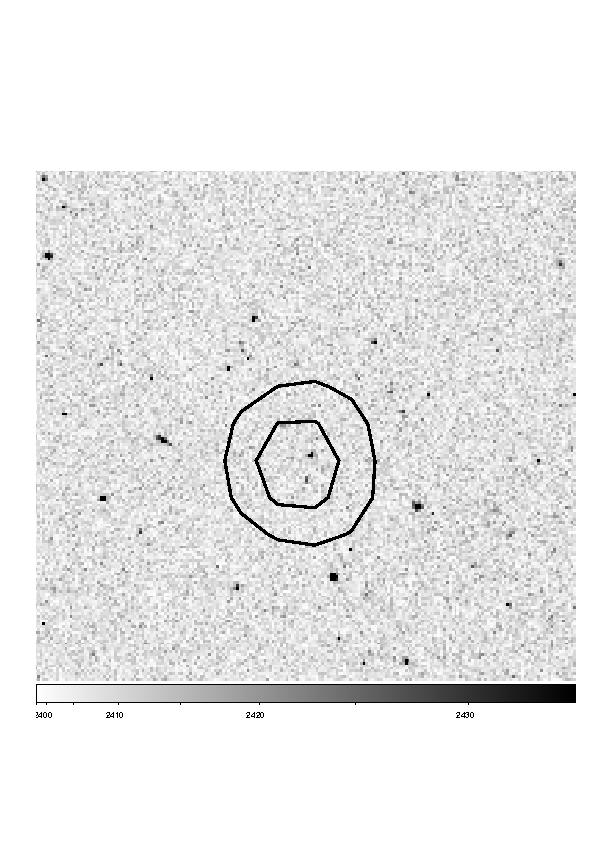}
\caption[]{$\chi^2$ image for Cl~J1202+4439 built from the SDSS u, g', r', i'
and z' images. The field is 3.7$\times$3.7 arcmin$^2$. Overlayed X-ray
contours are XMM data.}
\label{fig:cl28}
\end{figure}

\begin{figure}
\centering
\includegraphics[width=8cm,angle=0]{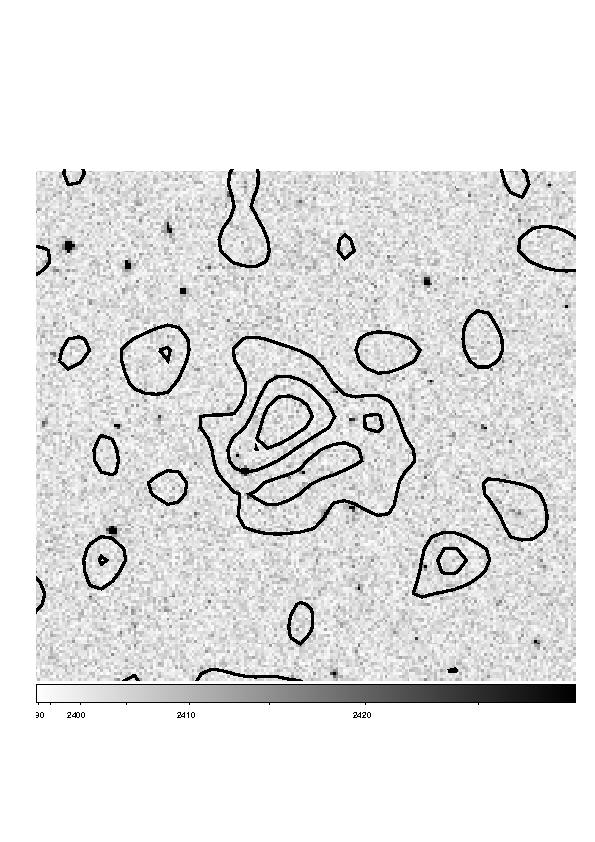}
\caption[]{$\chi^2$ image for Cl~J1207+4429 built from the SDSS u, g', r', i' and z' 
images. The field is 3.7$\times$3.7 arcmin$^2$.}
\label{fig:cl15}
\end{figure}

\clearpage

\begin{figure}
\centering
\includegraphics[width=8cm,angle=0]{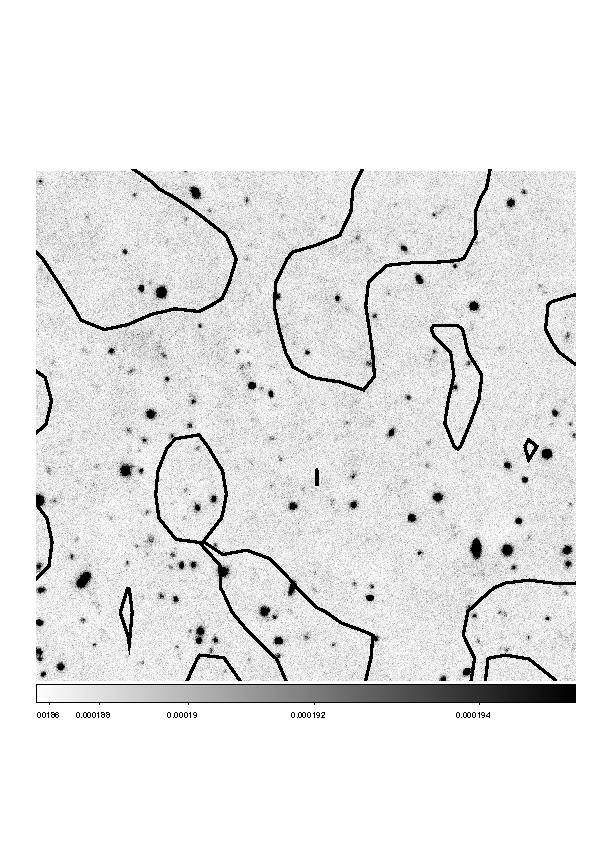}
\caption[]{i' image for Cl~J1213+3908 observed at ARC (completeness level:
i'$\sim$22.5).  The field is 3$\times$3 arcmin$^2$. }
\label{fig:cl27}
\end{figure}

\begin{figure}
\centering
\includegraphics[width=8cm,angle=0]{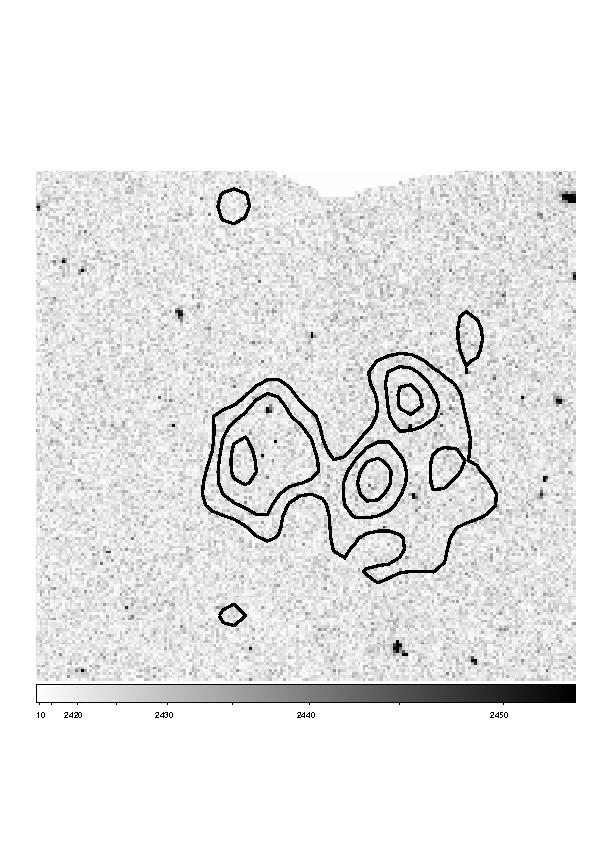}
\caption[]{$\chi^2$ image for Cl~J1213+3317 built from the SDSS u, g', r', i' and z' 
images. The field is 3.7$\times$3.7 arcmin$^2$. }
\label{fig:cl23}
\end{figure}

\begin{figure}
\centering
\includegraphics[width=8cm,angle=0]{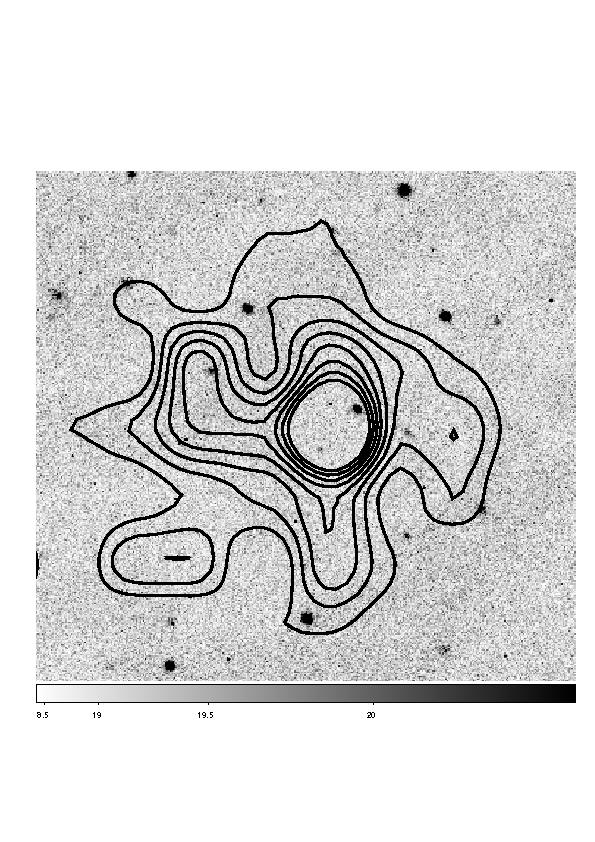}
\caption[]{I image for Cl~J1214+1254 observed at ESO (completeness level:
I$\sim$21). The field is 2$\times$2 arcmin$^2$.}
\label{fig:cl32}
\end{figure}

\begin{figure}
\centering
\includegraphics[width=8cm,angle=0]{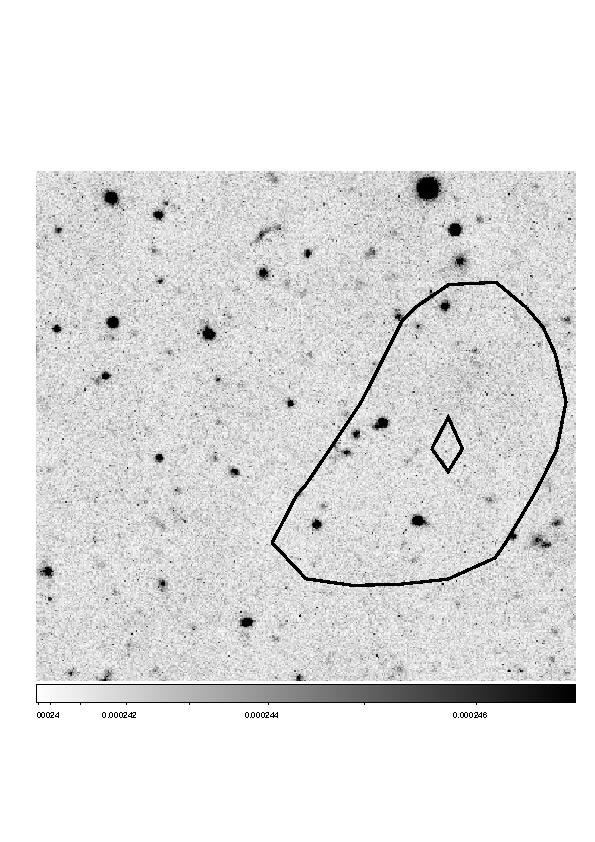}
\caption[]{i' image for Cl~J1216+3318 observed at ARC (completeness level:
i'$\sim$23.5).  The field is 1.8$\times$1.8 arcmin$^2$. }
\label{fig:cl36}
\end{figure}

\clearpage

\begin{figure}
\centering
\includegraphics[width=8cm,angle=0]{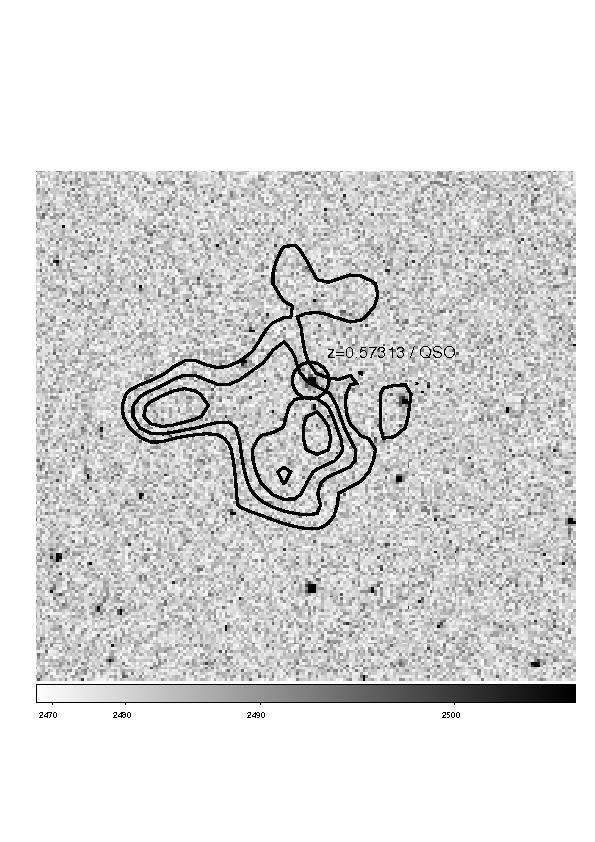}
\caption[]{$\chi^2$ image for Cl~J1234+3755 built from the SDSS u, g', r', i' and z' 
images. The field is 3.7$\times$3.7 arcmin$^2$.}
\label{fig:cl14}
\end{figure}

\begin{figure}
\centering
\includegraphics[width=8cm,angle=0]{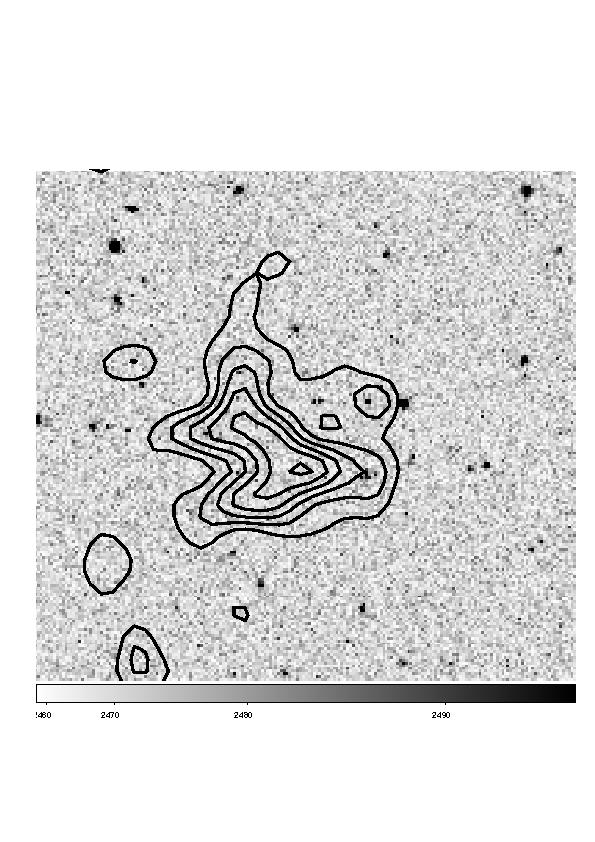}
\caption[]{$\chi^2$ image for Cl~J1237+2800 built from the SDSS u, g', r', i'
and z' images. The field is 3.7$\times$3.7 arcmin$^2$. }
\label{fig:cl31}
\end{figure}

\begin{figure}
\centering
\includegraphics[width=8cm,angle=0]{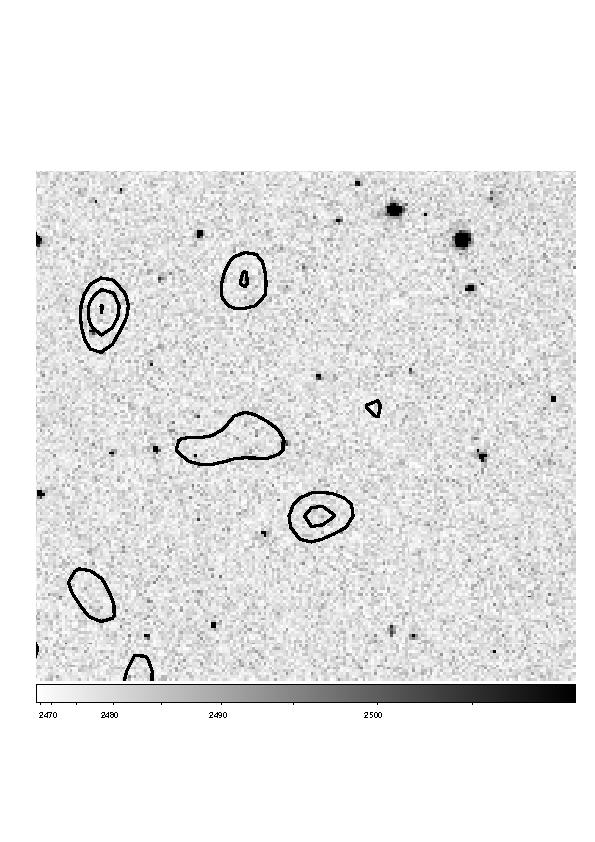}
\caption[]{$\chi^2$ image for Cl~J1259+2547 built from the SDSS u, g', r', i' and z' 
images. The field is 3.7$\times$3.7 arcmin$^2$.}
\label{fig:cl06}
\end{figure}

\begin{figure}
\centering
\includegraphics[width=8cm,angle=0]{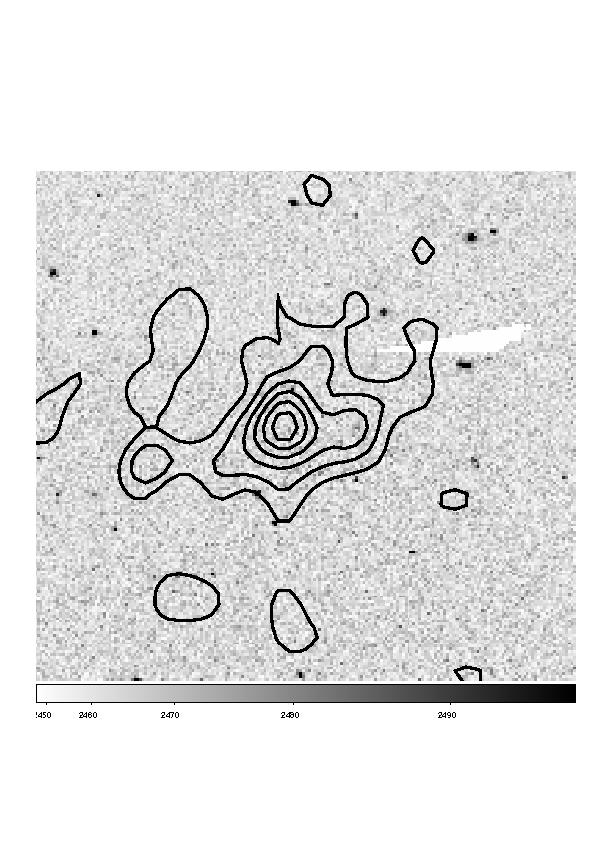}
\caption[]{$\chi^2$ image for Cl~J1343+2716 built from the SDSS u, g', r', i' and z' 
images. The field is 3.7$\times$3.7 arcmin$^2$.}
\label{fig:cl16}
\end{figure}

\clearpage

\begin{figure}
\centering
\includegraphics[width=8cm,angle=0]{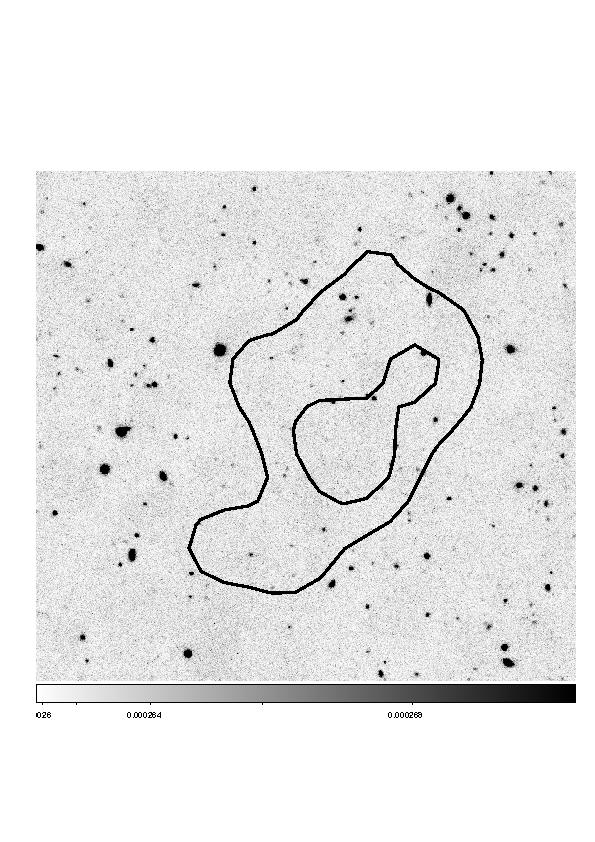}
\caption[]{i' image for Cl~J1350+6028 observed at ARC (completeness level:
i'$\sim$23).  The field is 3$\times$3 arcmin$^2$. }
\label{fig:cl29}
\end{figure}

\begin{figure}
\centering
\includegraphics[width=8cm,angle=0]{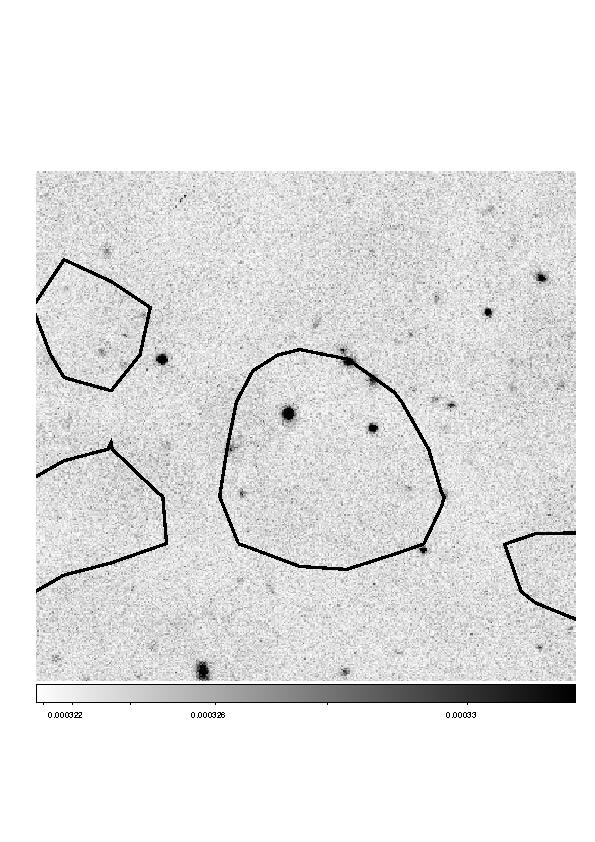}
\caption[]{i' image for Cl~J1411+5933 observed at ARC (completeness level:
i'$\sim$23.5).  The field is 1.8$\times$1.8 arcmin$^2$.}
\label{fig:cl24}
\end{figure}

\begin{figure}
\centering
\includegraphics[width=8cm,angle=0]{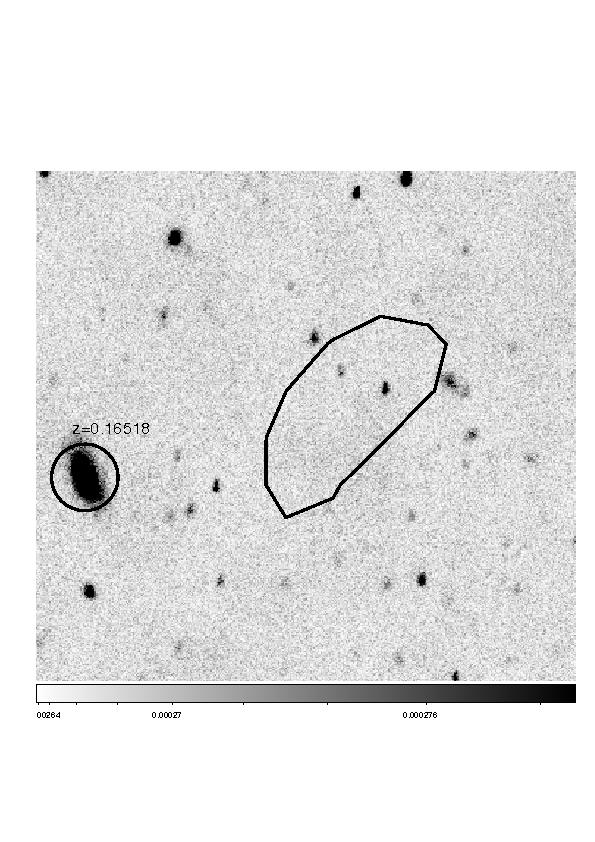}
\caption[]{$i'$ image for Cl~J1514+4351 observed at ARC (completeness level:
i'$\sim$22.5).  The field is 1.8$\times$1.8 arcmin$^2$.}
\label{fig:cl07}
\end{figure}

\begin{figure}
\centering
\includegraphics[width=8cm,angle=0]{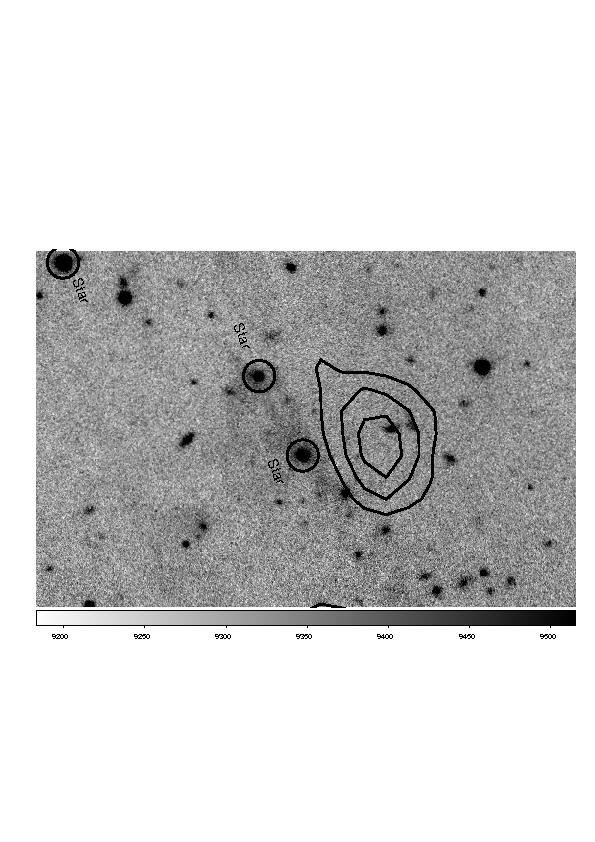}
\caption[]{r' image for Cl~J1651+6107 observed at Gemini (completeness level:
r'$\sim$24.5).  The field is 1.7$\times$1.1 arcmin$^2$. Overlayed
X-ray contours are XMM data. Circled objects are stars confirmed by
spectroscopy or image analysis.}
\label{fig:cl17}
\end{figure}

\end{appendix}


\begin{thebibliography}{}

\bibitem[]{} Adami C., Ulmer M.P., Romer A.K., et al., 2000, ApJS 131, 391

\bibitem[]{} Adami C., Picat J.P., Savine C., et al., 2006a, A\&A 451, 1159

\bibitem[]{} Adami C., Scheidegger R., Ulmer M.P., et al., 2006b, A\&A 459,
679

\bibitem[]{} Allen S., Schmidt R.W., Ebeling H., Fabian A.C., 
van Speybroeck L., 2004, MNRAS, 353, 457

\bibitem[]{} Andreon S., Willis J., Quintana H. et al., 2004, MNRAS, 353, 353

\bibitem[]{} Atanssova. M., Harvey J. E., 2003, SPIE, 5226, 275.

\bibitem[]{} Barkhouse W. A., Green P.J., Vikhlinin A., et al., 2006,
ApJ 645, 955

\bibitem[]{} Burke D., Collins C., Sharples R., et al., 1997, ApJ 488, L83

\bibitem[]{} Burrows C. J.,  Burg R. \& Giacconi R., 1992, ApJ, 392, 760

\bibitem[]{} Brodwin M., Brown M.J.I., Ashby M.L.N., et al., 2006, ApJ 651, 791

\bibitem[]{} Citterio O., Campana S., Conconi P., et al., 1999, SPIE, 3766,
198

\bibitem[]{} Dickey J.M., Lockman F.J., 1990, ARA$\&$A 28, 215

\bibitem[]{} Donahue M., Horner D.J., Cavagnolo K.W., Voit G.M., 2006, ApJ
  643, 730

\bibitem[]{} Ebeling H., Edge A., Bohringer H., et al., 1998, MNRAS 301, 881

\bibitem[]{} Fassbender R., Stegmaier J., Weijmans A.-M., et al., 2006,
SPIE 6266, 90

\bibitem[]{} Fukugita M., Shimasaku K. \& Ichikawa T., 1995, PASP 107, 945

\bibitem[]{} Harvey J. E., Atanssova M. \& Krywonos A. , 2004, SPIE, 5497, 636.

\bibitem[]{} Henry J.P., Gioia I.M., Maccacaro, T., et al., 1992, ApJ 386, 408

\bibitem[]{} Henry J.P., Gioia I.M., Mullis C., et al., 1997, AJ 114, 1293

\bibitem[]{} Holden B.P., Stanford S. A., Eisenhard P., Dickinson M., 2004, AJ, 127, 2484

\bibitem[]{} Jones L.R., Ponman T.J., Horton A., et al., 2003, MNRAS, 343, 627

\bibitem[]{} Lasker B.M., Sturch C.R., McLean B.J., et al., 1990, AJ 99, 2019

\bibitem[]{} Le F\`evre O., Saisse M., Mancini D., et al., 2003, SPIE 4841, 1670

\bibitem[]{} McLean B.J., Greene G.R., Lattanzi M.G., Pirenne B., 2000, ASPC
  216, 145

\bibitem[]{} Nichol R.C, Romer A.K., Holden B.P., et al., 1999, ApJL  521, L21

\bibitem[]{} Penprase B.E., Rhodes J.D., Harris E.L.,  2000, A\&A 364, 712

\bibitem[]{} Pierre M., Pacaud F., Duc P.A. 2006, MNRAS 372, 591

\bibitem[]{} Romer A., Nichol R., Holden B., et al., 2000, ApJS 126, 209

\bibitem[]{} Romer A., Viana P., Liddle A., Mann R., 2001, ApJ 547, 594

\bibitem[]{} Rosati P., Della Ceca R., Norman C., et al., 1998, ApJ 492, 21

\bibitem[]{} Sarazin C., 1986, Rev. Mod. Phys. 58, 1

\bibitem[]{} Scharf C., Jones L., Ebeling H., et al., 1997, ApJ 477, 79

\bibitem[]{} Schneider D.P., Hall P.B., Richards G.T., et al., 2005 130, 367

\bibitem[]{} Stanford S.A., Eisenhardt P.R., Brodwin M., et al., 2005,
ApJ 634, L129

\bibitem[]{} Szalay A.S., Connolly A.J., Szokoly G.P., 1999,
AJ 117, 68

\bibitem[]{} Ulmer M.P., Adami C., Covone G., et al., 2005, ApJ 624, 124

\bibitem[]{} Ulmer M.P., 1995, SPIE 2515, 280

\bibitem[]{} Vikhlinin A., McNamara B., Forman W., et al., 1998, ApJ 502, 558

\bibitem[]{} White R.L., Becker R.H.., Helfand D.J., Gregg M.D., 1997, ApJ 475, 479

\end{thebibliography}
\end{document}